\documentclass[aoas]{imsart}

\RequirePackage{amsthm,amsmath,amsfonts,amssymb}
\RequirePackage[authoryear]{natbib}
\RequirePackage[colorlinks,citecolor=blue,urlcolor=blue]{hyperref}
\RequirePackage{graphicx}

\usepackage{cleveref}
\usepackage[group-separator={,}]{siunitx}
\usepackage{psfrag,epsf}
\usepackage{enumerate}
\usepackage{apalike}
\usepackage{bm}
\usepackage{bbm}
\usepackage{url} 
\usepackage{alltt}
\usepackage[normalem]{ulem}
\usepackage[noend]{algpseudocode}
\usepackage{mathtools}
\usepackage[font=footnotesize]{subcaption}
\usepackage[font=footnotesize]{caption}
\usepackage{multicol}
\usepackage{color}
\usepackage{xcolor}
\usepackage[ruled,linesnumbered]{algorithm2e}
\usepackage{booktabs}
\usepackage{wrapfig}

\startlocaldefs
\makeatletter
\newcommand{\distas}[1]{\mathbin{\overset{#1}{\kern\z@\sim}}}%
\newsavebox{\mybox}\newsavebox{\mysim}
\newcommand{\distras}[1]{%
    \savebox{\mybox}{\hbox{\kern3pt$\scriptstyle#1$\kern3pt}}%
    \savebox{\mysim}{\hbox{$\sim$}}%
    \mathbin{\overset{#1}{\kern\z@\resizebox{\wd\mybox}{\ht\mysim}{$\sim$}}}%
}
\makeatother

\newcommand{\mbs}[1]{\boldsymbol{#1}}

\def\mnras{\it{MNRAS}}             
\def\apj{\it{ApJ}}                 
\def\apjs{\it{ApJS}}               
\def\prd{\it{Phys.~Rev.~D}}        

\IfFileExists{upquote.sty}{\usepackage{upquote}}{}

\algnewcommand{\LineComment}[1]{\State \(\triangleright\) #1}

\SetKwFor{For}{for }{ do}{}

\endlocaldefs

\begin{document}

\begin{frontmatter}
\title{Uncertainty Quantification of a Computer Model for Binary Black Hole Formation}

\begin{aug}
\author[A]{\fnms{Luyao} \snm{Lin}\ead[label=e1]{luyao\_lin\_2@sfu.ca}},

\author[A]{\fnms{Derek} \snm{Bingham}\ead[label=e2]{derek\_bingham@sfu.ca}},

\author[B]{\fnms{Floor} \snm{Broekgaarden}\ead[label=e3]{floor.broekgaarden@cfa.harvard.edu}}

\and
\author[C,D,E]{\fnms{Ilya} \snm{Mandel}\ead[label=e4]{ilya.mandel@monash.edu}}
\address[A]{Simon Fraser University,
\printead{e1,e2}}

\address[B]{Harvard-Smithsonian Center for Astrophysics,
\printead{e3}}

\address[C]{School of Physics and Astronomy, Monash University, Clayton, Victoria 3800, Australia, \printead{e4}}

\address[D]{The ARC Center of Excellence for Gravitational Wave Discovery -- OzGrav, Australia}

\address[E]{Birmingham Institute for Gravitational Wave Astronomy and School of Physics and Astronomy, University of Birmingham, Birmingham, B15 2TT, United Kingdom}

\end{aug}

\begin{abstract}
In this paper, a fast and parallelizable method based on Gaussian Processes (GPs) is introduced to emulate computer models that simulate the formation of binary black holes (BBHs) through the evolution of pairs of massive stars. Two obstacles that arise in this application are the {\it{a priori}} unknown conditions of BBH formation and the large scale of the simulation data. We address them by proposing a local emulator which combines a GP classifier and a GP regression model. The resulting emulator can also be utilized in planning future computer simulations through a proposed criterion for sequential design. By propagating uncertainties of simulation input through the emulator, we are able to obtain the distribution of BBH properties under the distribution of physical parameters.
\end{abstract}


\end{frontmatter}


\section{Introduction}\label{sec:intro}
Scientists frequently explore complex phenomena by means of computer models that simulate the behavior of these systems. In some cases, the CPU time required to evaluate the model can take hours to months \citep[e.g.][]{gramacy2008bayesian}, while in others the model may be fast to evaluate on a super computer, but is not readily available to those who need it \citep[e.g.][]{kaufman2011efficient,lawrence2017mira}. In either case, Gaussian process emulators \citep{sacks1989design} are often used to stand in for the computer model (or simulator).  In this paper, we propose a new type of emulator where a large simulation design is available, but there are unknown constraints that specify where a model output will occur.

\begin{wrapfigure}{r}{0.4\textwidth}
  \begin{center}
    \includegraphics[width=0.37\textwidth]{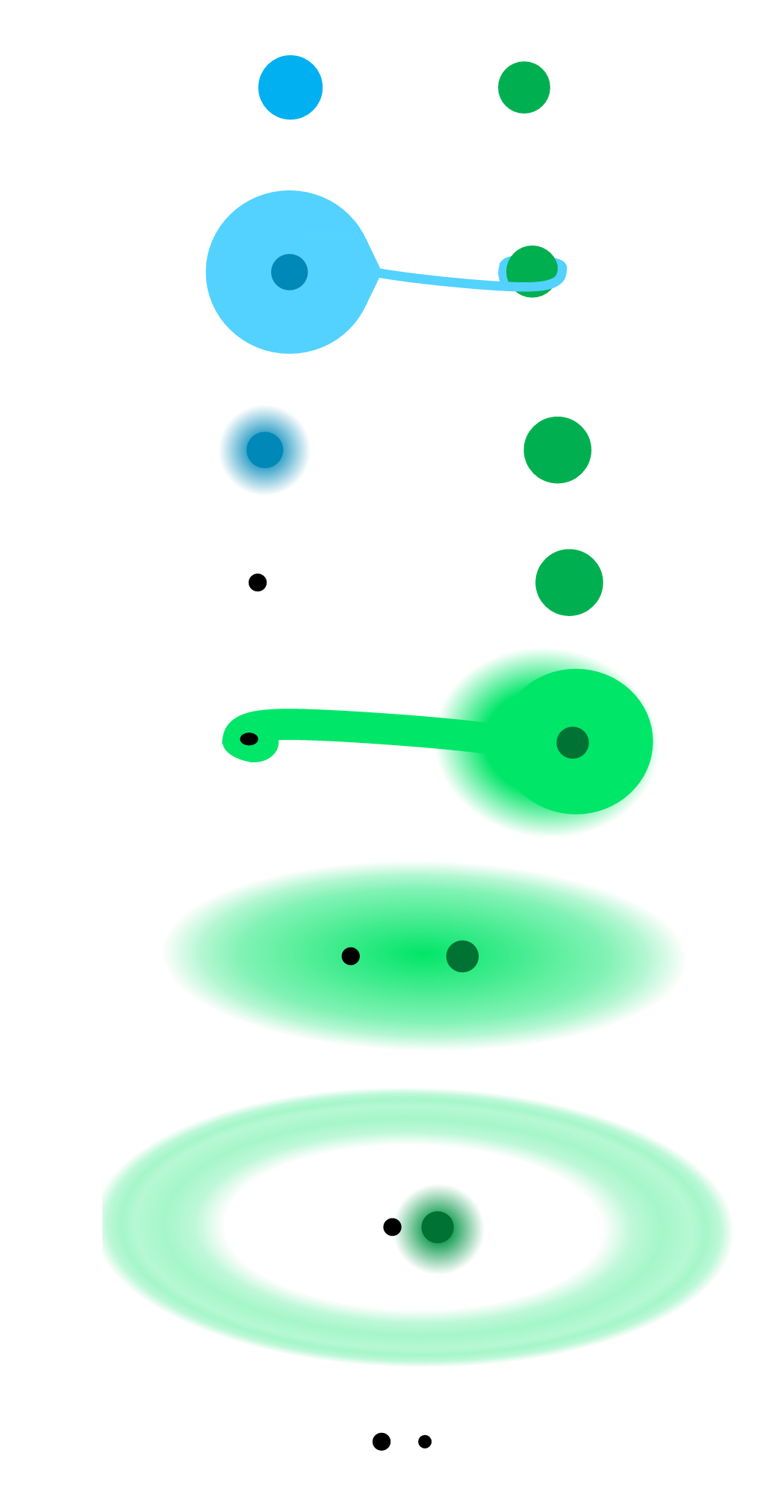}
  \end{center}
  \caption{Schematic view of the formation pathway of a BBH system through classical isolated binary evolution via the common-envelope phase. Panels from top to bottom show: initial stellar binary; mass transfer from the evolving and expanding more massive star (blue) onto its less massive companion (green); continued evolution of initially more massive star until it collapses into a black hole; mass transfer from the initially less massive star onto the black hole which leads to the formation of a common envelope of gas, significant drag, and rapid spiral-in; the ejection of the common envelope leaving behind a tighter binary; and the collapse of the companion into a black hole.  Adapted from \citet{MandelFarmer:2018}.}\label{fig:BBH_scenario}
\end{wrapfigure}

The application that motivated the proposed methodology was to construct a fast emulator for binary population synthesis simulation codes that study characteristics of binary black hole (BBH) mergers. Population synthesis codes typically begin with a binary star system at birth, and determine the system's evolutionary outcome. Figure \ref{fig:BBH_scenario} depicts an example of the binary evolution pathway modelled by such simulation codes. These computer models are quite fast, potentially yielding millions of model evaluations per day. However, given the high dimensionality of the input and complexity of binary stellar evolution, in practice many billions of binaries need to be simulated to perform an experiment that is sufficiently large to make scientific inferences. This can amount to computing times of years.

To address this issue, present-day simulation studies in these settings make compromises such as sacrificing accuracy for speed by adopting approximate/simplifying algorithms, or restricting the exploration of physical assumptions to limit the number of simulations. Here, by introducing a fast statistical surrogate model for the simulation codes, we aim to preserve both accuracy (with uncertainty measures) and free exploration of physical parameters.

The first challenge facing the emulation of population synthesis is the large scale of these experiments, which renders traditional GPs prohibitive due to computational limitations resulting from the inversion of the covariance matrix in the Gaussian likelihood \citep[see e.g.][]{kaufman2011efficient,2017IAUS..325...46B}. There are several approaches dealing with GP modeling with large-scale data. Innovations to the full covariance matrix are often proposed to alleviate the computational burden \citep[e.g.][]{kaufman2011efficient,cressie2008fixed,quinonero2005unifying}.
\citet{gramacy2015local} side-step the issue by using smaller, local GPs (laGP). Inspired by the laGP, the approach proposed herein focuses on building local models by choosing informative neighboring designs for input locations of interest.

A second challenge in population synthesis emulation is that the success rate of producing a BBH via sampling from the input distribution of initial conditions is extremely low. Typically, one BBH is formed per $\sim\ 10^3 - 10^6$ binaries simulated \citep[e.g.][]{2002ApJ...572..407B,2018PhRvD..98h3017T,2018MNRAS.481.1908K,broekgaarden2019stroopwafel}, so the vast majority of computational time is spent on simulations that do not produce an outcome. Improvements to the success rate have been proposed through Adaptive Importance Sampling (AIS) \citep{broekgaarden2019stroopwafel}, or Markov Chain Monte Carlo (MCMC) \citep{2018ApJS..237....1A}, but in both cases it is challenging to scale to high dimensions and non-trivial to perform inference.

Lastly, since the input conditions that allow for BBH formation are not known in advance, it is desirable that the surrogate model addresses this issue. Latent variable models have been employed to provide a probabilistic quantification of the unknown constraints.  For example, \citet{gramacy2010optimization} and \citet{gelbart2014bayesian} used a GP classifier in an attempt to optimize a system in the presence of unknown constraints. In our work, we take a Bayesian approach to make inferences for the latent variable model. However, when it comes to large data sets, fully Bayesian inference in our setting becomes computationally intensive as all components in the Markov Chain need to be updated and stored. 

In this paper, new methodology is proposed for large-sample emulation for computer models with unknown constraints.  The proposed emulator combines a fast-to-compute GP classifier and a local GP model to provide predictions with uncertainty quantification for population synthesis codes.  

The  paper is organized as follows. Section \ref{sec:BBH} details the  population synthesis simulation model that motivated this work, namely the COMPAS model\footnote{Compact Object Mergers: Population Astrophysics and Statistics (COMPAS, \href{https://compas.science}\url{https://compas.science}): a platform for the exploration and study of populations of compact binaries formed through isolated binary evolution.}.   In Section \ref{sec:emu_model}, new methodology that combines a local GP classifier and a local GP emulator is proposed, followed by the introduction of a sequential design criterion for improving the emulator. Synthetic examples illustrating the proposed method are given in Section \ref{sec:emu_example}, and emulation results for BBH formation are presented in Section \ref{sec:COMPAS_res}. The proposed methodology is fairly general and can easily be adapted to other population synthesis codes or models with unknown constraints.  The paper concludes with comments and future work.

\section{Population synthesis of BBH mergers}\label{sec:BBH}
The COMPAS model that motivated this work is a rapid binary population synthesis code that focuses on gravitational-wave astrophysics. In particular, it is designed to study uncertainties in binary evolution and to optimize the information that can be obtained from simulations \citep{stevenson2017formation,2018MNRAS.477.4685B,2018MNRAS.481.4009V,Neijssel:2019}.

\begin{table}[!t]
    \small
    \caption{Input/Output of COMPAS models}\label{tab:data}
    \centering
    \begin{tabular}{p{7.5cm}rr}
    \hline
    \textbf{Input} &    \textbf{Range} & \textbf{\textsl{Distribution}}\\
    \hline
    \textbf{Initial conditions: ${x}$} & & \\
    $m_1$ : the mass of the initially more massive star & $[8,150]\, \rm{M}_{\odot}$  & Power law(-2.35)\\
    $m_2$:  the mass of the initially less massive star  & $(0.1\, \rm{M}_{\odot}, m_1]$ & Uniform \\
    $a$: the initial orbital separation & $[0.01,1000] \, \rm{AU}$  & Power law(-1)\\
    $\mathbf{v_{\rm{i}}}$ : supernova natal kick vector for supernova $i$, $i=1,2$, &   &    \\
    including: & & \\
    \hspace{.5cm}$v_{\rm{i}}$ -  magnitude of the supernova natal kick ($\mathrm{km}\, \mathrm{s}^{-1}$) &   $[0,\infty)$ &  Maxwellian  \\
    \hspace{.5cm}$\theta_{\rm{i}}$ - polar angle defining the direction of the natal kick &$[0,\pi]$ & Uniform\\
    \hspace{.5cm}$\phi_{\rm{i}}$ - azimuthal angle defining the direction of the natal kick  &$[0,2\pi]$ & Uniform\\
    \hspace{.5cm}$\omega_{i}$ - mean anomaly & $[0,2\pi]$ & Uniform\\
    \textbf{Hyperparameters: ${t}$} & & \\
    $Z$: the metallicity & $[0.0001, 0.03]$ &  \\ 
    $\alpha$ : the common envelope efficiency parameter & $[0,10]$ & \\

    $\sigma$ : 1D root-mean-square value representing a typical supernova kick & $[0,1000]$ $\rm{km}$ {s}$^{-1}$ &  \\
    flbv: multiplication factor for the mass loss rate during the luminous blue variable (LBV) phase & $[0,10]$ & \\
    \hline
    \textbf{Output} & & \\
    \hline
    Success: indicates whether BBH is formed & $\{0,1\}$ & \\
    $\mathcal{M}_c$: chirp mass of BBH & $(0,150)\ \rm{M}_\odot$ or NA & \\
    \hline
    \end{tabular}
\end{table}

There are two types of input in the COMPAS model. One is the set of initial conditions intrinsic to each binary, denoted by ${x}$. These provide the state of the binary at formation (top of Figure \ref{fig:BBH_scenario}), such as the initial stellar masses and the distance between the stars. Another type of input is the set of population hyper-parameters, denoted by ${t}$, which is shared between all binaries in a population. These hyper-parameters can be thought of as parametrizations of the differential equations governing stellar and binary evolution. The true values of the hyper-parameters ${t}$ are unknown in advance. An important goal (not addressed here) is to infer their values by comparing the BBH properties predicted by simulations with different choices of ${t}$ with observations \citep{Mandel2017GravitationalWS}.

A key output for BBH population synthesis is the chirp mass of the BBH, which is a combination of the masses that are typically best measured from the gravitational-wave signal \citep{1963PhRv..131..435P}. For the purpose of studying BBH formation, the COMPAS output is summarized by the chirp mass of the BBH, or if the binary does not evolve into a BBH, the output is ``NA''.  A relevant list of COMPAS inputs is provided in Table \ref{tab:data}. The initial conditions of a binary follow observationally constrained distributions; simplified versions of these are specified in the third column of the table.  A key challenge is that the regions of input space that result in ``NA'' outputs are unknown - thus the unknown constraints.

The COMPAS model, like many other binary population synthesis codes, has a relatively high-dimensional input and, more importantly, a low success rate for BBH formation. Specifically, the  COMPAS model used here requires a 15-dimensional input, and produces a success rate that is below $1\%$ \citep{broekgaarden2019stroopwafel}. In the next section, we introduce an emulator for COMPAS that addresses the large sample size ($>10^6$) and also the unknown constraints that result in the low success rate for BBH formation in the simulations.

\section{Methodology: local surrogate models with unknown constraints}\label{sec:emu_model}

In this section, new methodology for emulating large-sample computer experiments on simulators with unknown constraints is proposed.  Before introducing the components of the approach, some notation is first introduced.

Denote the deterministic computer model as $m(\cdot)$ with inputs $x$ that, without loss of generality, belong to the $d$-dimensional unit cube. The unknown constraints define the subset of input space, $C$, where the simulator returns a univariate, real-valued response ({i.e.,  the \em constraint region}).  For the applications we consider, $C$ is assumed to be the union of non-overlapping, compact regions of the input space.

An indicator function, $y(x)$, is used to identify whether or not an input results in a real-valued output.  That is, $y(x)=1$ if $\{x\in C\}$ and zero otherwise. The computer model output can then be expressed as
\begin{equation}
m({x}) = \left\{\begin{array}{ll} \textrm{NA}\ \ \ \ \ \ \ \  & \textrm{if } y(x)=0\\
z(x)\ \ \ \ \ \ \ \  & \textrm{if } y(x) = 1\end{array}\right.,
\label{eq:decomp}
\end{equation}
where $m(x)=\textrm{NA}$ corresponds to $x\not\in C$, and $z({x})$ is the real-valued computer model response for $\{x\in C\}$. 

Let $\boldsymbol{X}=(x_1,x_2,\ldots,x_N)'$ be the $N$-run computer experiment design matrix, and  $\mbs{m}=(m_1,m_2,\ldots,m_N)'$ be the model outputs. The corresponding indicator labels are denoted $\boldsymbol{y} = (y_1,y_2,\ldots,y_N)'$. We partition the design, matrix $\mbs{X}$, into the {\em active set}, $\{x:x\in\mbs{X} \textrm{ and } x\in C\}$, and the  {\em null set},  $\{x:x\in\mbs{X} \textrm{ and } x\not \in C\}$.

As we shall see, the formulation in (\ref{eq:decomp}) will allow us to (i) obtain a probabilistic representation of the unknown constraints through $y({x})$, and (ii) decompose prediction uncertainties into components that correspond to unknown constraints and emulation errors, respectively. 
In the COMPAS model, for example, the probability of producing a BBH at an unsampled input, ${x}^*$, can be quantified by ${P}(y({x}^*)=1)$. This is useful since one can, for example, learn about the initial conditions that are likely to lead to BBH formation conditional on the observed simulator responses.  Of course, one is also interested in predicting the chirp mass, and thus we are also interested in estimating $m({x}^*)$.

In the applications considered, $N$ is large, and the computation involved in fitting a conventional GP is prohibitive. A two-step procedure is proposed  to address the big $N$ problem in the presence of unknown constraints.  We take a similar approach as \citet{gramacy2010optimization} where a classifier is used to identify whether or not an input is in the constraint region and an independent GP is used to emulate $z(x)$. First, the indicator function, or {\em constraint function}, is modelled  using  a local GP classifier. Second, when ${y}({x}^*)$ is predicted to be one, a local GP emulator is constructed to predict the simulator output ${m}({x}^*)$. If $y({x}^*)$ is predicted to be zero, ${m}({x}^*)$ is predicted as NA. 

The proposed local classification model is presented in Section \ref{subsec:classification}, followed by a local response surface model in Section \ref{subsec:emulation}. A holistic view of the procedure is given in Section \ref{subsec:local_emu}. Finally, we propose new sequential design methodology in Section \ref{subsec:sequential} for the selection of new simulation trials.  

\subsection{Classification}\label{subsec:classification}

A logistic GP classifier \citep{williams2006gaussian} is used to model the constraint function.  Specifically,
\begin{eqnarray}
P(y_i = 1) \equiv q_i & = & (1+e^{-f_i})^{-1},\ \ \  \mbox{  or equivalently, }\ \ \ f_i = \log\frac{P(y_i = 1)}{1-P(y_i = 1)},
\end{eqnarray}
where $\boldsymbol{f}=(f_1,f_2,\ldots,f_N)'$ is a vector of latent variables describing the log-odds of $\{y_i=1\}$.

A mean-zero GP is used to model the latent variables $\boldsymbol{f}$ as a function of the inputs $\boldsymbol{X}$. That is,
\begin{equation}\label{eq:MVN}
\boldsymbol{f}\sim \mathcal{N}\left(\boldsymbol{0}, \boldsymbol{\Sigma}_f \right),
\end{equation}
where  $\boldsymbol{\Sigma}_f$ is the $N\times N$ covariance matrix with elements determined by a stationary covariance function. Throughout, we use the squared-exponential covariance function \citep{sacks1989design}
\begin{equation}
\textrm{cov}(f_i,f_j) = \eta^{-1} \cdot \exp\left(-\sum_{l=1}^d \frac{\left(x_{i,l}-x_{j,l}\right)^2}{\phi_{l}^2} \right),
\end{equation}
where $\eta$ denotes the precision parameter and $\phi_{l}$ ($\phi_{l}>0$)  is the length-scale parameter for the $l$-th dimension.

When $N$ is large, the time required to evaluate the Gaussian likelihood due to inverting the covariance matrix can be exceedingly long. To address this, alternate methods have been proposed to alleviate the computational burden by imposing simplifying assumptions on the covariance matrix \citep[e.g.][]{kaufman2011efficient,cressie2008fixed,quinonero2005unifying}. Another approach is to construct smaller, local designs in the neighborhood of the unsampled input, $x^*$, to emulate $m(x^*)$ \citep{gramacy2015local}. Here, we propose to use some of the elements outlined in \citet{gramacy2015local} for inference on the models presented in Sections \ref{subsec:classification} and \ref{subsec:emulation}. For classification, we use the $n$ nearest neighbors ($n\ll N$) to the input of interest, $x^*$. Denote the inputs, with their outputs, closest to $x^*$ as $B(x^*)=(\mbs{X}^b,\mbs{m}^b)$. The computation for emulating $y(x^*)$ can then be reduced from $O(N^3)$ to $O(n^3)$. Let $\mbs{y}^b$ and $\mbs{f}^b$ denote the class label and latent log-odds variable at $\mbs{X}^b$. The joint distribution of $\mbs{y}^b$ and $\mbs{f}^b$ given the local GP classifier parameters, $\mbs{\phi}^b$ and $\eta^b$, is
\begin{eqnarray*}
&   & p(\mbs{y}^b,\mbs{f}^b|\boldsymbol{\phi}^b,\eta^b)\\
& = & p(\mbs{y}^b|\mbs{f}^b)\cdot p(\mbs{f}^b|\mbs{\phi}^b,\eta^b) \\ 
& = & \left[\prod_{i=1}^n \bigg(\frac{e^{f_i^b}}{1+e^{f_i^b}}\bigg)^{y_i^b}  \bigg(\frac{1}{1+e^{f_i^b}}\bigg)^{1 - y_i^b} \right] \cdot \frac{1}{\sqrt{2\pi}^n|\mbs{\Sigma}_f^b|^{1/2}}\cdot \exp \bigg[-\frac{1}{2}(\mbs{f}^{b})^T(\mbs{\Sigma}_f^b)^{-1}(\mbs{f}^b)\bigg],
\end{eqnarray*}
where $\mbs{\Sigma}_f^b$ is the $n\times n$ covariance matrix of $\mbs{f}^b$, $p(\mbs{y}^b|\mbs{f}^b)$ is joint probability mass function for $\mbs{y}^b$ (i.e., independent Bernoulli random variables), and $p(\mbs{f}^b|\mbs{\phi}^b,\eta^b)$ is a local GP in the form of (\ref{eq:MVN}).  The joint posterior distribution of $\mbs{f}^b$, $\mbs{\phi}^b$ and $\eta^b$ can be expressed as
\begin{eqnarray}\label{eq:GPC_posterior}
p(\mbs{f}^b,\mbs{\phi}^b,\eta^b|\mbs{y}^b)& = & \frac{p(\mbs{f}^b,\mbs{\phi}^b,\eta^b,\mbs{y}^b)}{p(\mbs{y}^b)}\\
& \propto & p(\mbs{y}^b,\mbs{f}^b|\boldsymbol{\phi}^b,\eta^b) \cdot \pi(\mbs{\phi}^b,\eta^b),\nonumber
\end{eqnarray}
where $\pi(\mbs{\phi}^b,\eta^b)$ is the joint prior distribution for $\mbs{\phi}^b$ and $\eta^b$. 

To draw posterior samples of $\mbs{f}^b$, $\mbs{\phi}^b$ and $\eta^b$, 
single site Metropolis-Hastings (MH) MCMC can be employed \citep{hastings1970monte}. However, since ${f_i}^b$'s are correlated, independent sampling is inefficient. Instead, elliptical slice sampling \citep{murray2010elliptical} is adopted for $\mbs{f}^{b}$, with MH steps used for $\mbs{\phi}^b$ and $\eta^b$. {Elliptical slice sampling is practical in this setting because (a) it has a 100\% acceptance rate, (b) the simultaneous update of vector $\mbs{f}^b$ and (c) there are no algorithm parameters to tune.} Details of the sampling procedure can be found in Appendix \ref{app:MCMC}.

Once posterior samples of $\mbs{f}^{b}$, $\mbs{\phi}^b$ and $\eta^b$ are obtained, the latent log-odds variable at $x^*$, denoted by $f(x^*)$, can be predicted with
\begin{equation}\label{eq:logodds_post}
    [f(x^*)|\mbs{f}^{b},\mbs{\phi}^b,\eta^b] \sim \mathcal{N} \Big[\mbs{r}^T(x^*)(\mbs{\Sigma}_f^b)^{-1}\mbs{f}^{b},1/\eta^b-\mbs{r}^T(x^*)(\mbs{\Sigma}_f^b)^{-1}\mbs{r}(x^*)\Big],
\end{equation}
where $\mbs{r}(x^*)$ is the the vector of covariances between $f(x^*)$ and $\mbs{f}^{b}$.

To summarize, sampling of $f(x^*)$ given $B(x^*)$ starts with drawing a posterior sample of $(\mbs{f}^{b},\mbs{\phi}^b,\eta^b)$, with which the posterior mean and variance in \eqref{eq:logodds_post} are computed. The procedure is concluded by drawing a random sample from the resulting Gaussian process.

\begin{algorithm}[!h]
 \KwIn{model input of interest $x^*$, local design size $n$, simulation data $D = (\mbs{X},\mbs{m})'$} 
 \KwOut{posterior sample of ${y}(x^*)$ and log-odds $f(x^*)$}
 $\mbs{x}^{b} \leftarrow n$ neighboring points to $x^*$ in the simulation data\;
 $\mbs{y}^{b} \leftarrow$ class labels for $\mbs{X}^{b}$ \;
 \eIf{$\mbs{X}^{b}$ all in null set, i.e., $\mbs{y}^{b}=\mbs{0}$}{
 predict ${y}(x^*)$ to be $0$ with probability $1$. 
 }{
 \eIf{$\mbs{X}^{b}$ all in active set, i.e., $\mbs{y}^{b}=\mbs{1}$}{
 predict ${y}(x^*)$ to be $1$ with probability $1$.
 }{ build GP classifier with $\mbs{X}^{b}$ and $\mbs{y}^{b}$ \;
    predict ${y}(x^*)$ using the posterior distribution of the classification model parameters and $\mbs{f}^b$. 
  }
 }
\caption{Classify $y(x^*)$ with local GP classifier}
\label{ag:y}
\end{algorithm}

A benefit of using $B(x^*)$ instead of the full simulation data, aside from reduced computational complexity in the GP classifier, is the opportunity to forego the MCMC step altogether when all constraint function labels are the same in $B(x^*)$. As illustrated in Algorithm \ref{ag:y}, we start with the full simulation data, an input of interest $x^*$ and a user-specified local design of size $n$. An $n$ nearest neighbor design, $\mbs{X}^{b}$, is then constructed. If the corresponding class labels $\mbs{y}^b$ are identical, $y(x^*)$ can be simply set to the common value of $\mbs{y}^b$ with probability 1. In cases where the class labels, $\mbs{y}^b$, of the local design are not identical, the aforementioned MCMC procedure is carried out to make inferences about model parameters and $y(x^*)$. {The choice of $n$ involves a trade-off between computational burden and classification accuracy. For the classification problem, setting $n$ as large as permitted by computing resources can help with prediction accuracy. On the other hand, the evaluation of the log-likelihood within the MCMC grows at $O(n^3)$, and there will be more latent variables to be sampled with larger $n$, thereby  resulting in a slower MCMC runs. In the BBH application where emulation speed is important, we found $n=50$ to be a satisfactory after trying different choices for $n$ and examining the corresponding classification accuracy and runtime.} The choice of the distance metric can be made based on the application of interest, and is discussed later in Section \ref{subsec:COMPAS_emu}.

\subsection{Response surface model}\label{subsec:emulation}

We propose to use a local GP emulator for the response surface $z(x)$ on the active set. Let $\mbs{z}^a = (z_1^a,z_2^a,\ldots,z^a_{N^a})'$ denote the subset of simulation outputs $\mbs{m}$ that are real-valued, with corresponding design points $\mbs{X}^a = (x^a_1,x^a_2,\ldots,x^a_{N^a})'$. 

Generally speaking, we can emulate $z(x)$ with $\mbs{X}^a$ and $\mbs{z}^a$ by assuming a constant mean GP model, 
$$\mbs{z}^a \sim \mathcal{N}({\mu} \cdot  \mbs{1_{N^a}},\mbs{\Sigma}_z),$$
where ${\mu}$ is the constant mean and $\mbs{\Sigma}_z$ is the covariance matrix. We adopt the squared-exponential covariance function for $\mbs{\Sigma}_z$, with
\begin{equation}
\textrm{cov}(z^a_i,z^a_j) = \lambda^{-1} \cdot \exp\left(-\sum_{l=1}^d \frac{\left(x^a_{i,l}-x^a_{j,l}\right)^2}{\psi_{l}^2} \right),
\end{equation}
where $\lambda$ is the precision parameter, $\psi_{l}$ ($\psi_{l}>0$)  is the length-scale parameter for the $l$-th input dimension, and $\mbs{\psi} = (\psi_1,\psi_2,\ldots,\psi_d)'$.

\begin{algorithm}[!t]
 \KwIn{model input $x^*$, initial neighboring design $B(x^*)=(\mbs{X}^{b},\mbs{m}^{b})$, maximum neighbor size $n^M$, training data $D$} 
 \KwOut{predictive distribution of ${z}(x^*)$}
 $A(x^*) \leftarrow $ active points in $B(x^*)$ \;
 $T \leftarrow $ points in $D\backslash A(x^*)$ that are the first nearest neighbors of $A(x^*)$\;
 $T^a \leftarrow $ active simulation data in $T$\;
 \While{$ T^a \neq \varnothing $ \textbf{and} size of $A(x^*)$ is less than $n^M$}{
 Add $ T^a $ to $A(x^*)$\\
 Update $T$ and $T^a$
 }
 \tcp*[r]{Expand neighboring points}
\eIf{size of $A(x^*)$ is larger than $n^M$}{Remove from $A(x^*)$ simulations with inputs that are furthest from $x^*$ to maintain design size of $n^M$}{Keep $A(x^*)$ as the final local design}\;
Build local approximate GP with $A(x^*)=(\mbs{X}^{ab},\mbs{z}^{ab})$ \;
Obtain predictive distribution of $z(x^*)$
\caption{Emulating $z(x^*)$ with local approximate GP}
\label{ag:z}
\end{algorithm}

For the reasons discussed in Section \ref{subsec:classification}, evaluation of the Gaussian likelihood becomes infeasible with large $N^a$. To address this, a local GP emulator is adopted instead of the global GP stated above. For an unsampled input of interest, $x^*$, a local design consisting of only active simulations is constructed to emulate $z(x^*)$. One might be tempted to create this local design with nearest neighbors to $x^*$ from the active set. However, some of these active neighbors might come from different compact subsets of the constraint region than that of $x^*$, resulting in inclusion of simulation data that can represent very different behavior than the neighborhood of $z(x^*)$. To construct this active local design, the set of active simulations in local data $B(x^*)$, denoted by $A(x^*)$, is used as the starting point in an iterative algorithm. Recall that $D=(\mbs{X},\mbs{m})$ represents the full simulation data. For each iteration, we first search the remaining simulations for ones with inputs that are the nearest neighbors of the design points in $A(x^*)$, and denote these simulation data by $T = (\mbs{X}^\textrm{t},\mbs{m}^\textrm{t})$. Next, the active simulations in $T$ are included in $A(x^*)$. The search continues until either the maximum design size $n^M$ for $A(x^*)$ is reached or all first nearest neighbors to $A(x^*)$ are from the null set. If the algorithm ended up with more than $n^M$ simulations after the last iteration, observations with inputs that are furthest from $x^*$ will be removed. Denote the resulting local data with $n^{ab}$ observations as $A(x^*)=(\mbs{X}^{ab},\mbs{z}^{ab})'$. This procedure is summarized in Algorithm \ref{ag:z}.

The GP likelihood is
\begin{equation}
\label{eq:lik_Z}
p(\mbs{z}^{ab}|\mbs{\psi}^b,\lambda^b,{\mu}^b) = \frac{1}{\sqrt{2\pi}^{n^{ab}}|\mbs{\Sigma}_z^b|^{1/2}}\exp\left[-\frac{1}{2}(\mbs{z}^{ab}-{\mu}^b\cdot\mbs{1})'(\mbs{\Sigma}_z^b)^{-1}(\mbs{z}^{ab}-{\mu}^b\cdot\mbs{1})\right],
\end{equation}
where $\mbs{\Sigma}_z^b$ is the covariance matrix for $\mbs{z}^{ab}$.
The posterior of $(\mbs{\psi}^b,\lambda^b,\mu^b)$ is then
\begin{align}\label{eq:post_GP_par}
p(\mbs{\psi}^b,\lambda^b,\mu^b|\mbs{z}^{ab}) & = \frac{p(\mbs{\psi}^b,\lambda^b,\mu^b,\mbs{z}^{ab})}{p(\mbs{z}^{ab})}\nonumber \\
& \propto p(\mbs{z}^{ab}|\mbs{\psi}^b,\mu^b,\lambda^b)\cdot \pi(\mbs{\psi}^b,\mu^b,\lambda^b),
\end{align}
where $\pi(\mbs{\psi}^b,\mu^b,\lambda^b)$ is the prior distribution of the parameters. 

The conditional distribution of ${z}(x^*)$ is
\begin{equation}\label{eq:Z_post}
    [{z}(x^*)|\mbs{z}^{ab},\mbs{\psi}^b,\lambda^b,\mu^b] \sim \mathcal{N} \Big(\mbs{k}'(x^*)(\mbs{\Sigma}^b_z)^{-1}(\mbs{z}^{ab}-\mu^b\mbs{1})+\mu^b\mbs{1},(\lambda^b)^{-1}-\mbs{k}'(x^*)(\mbs{\Sigma}^b_z)^{-1}\mbs{k}(x^*)\Big),
\end{equation}
where $\mbs{k}(x^*)$ represents the covariance between $z(x^*)$ and $\mbs{z}^{ab}$. To sample $z(x^*)$ given $A(x^*)$, we first draw samples of $(\mbs{\psi}^b,\lambda^b,\mu^b)$ from the posterior distribution in \eqref{eq:post_GP_par} (with standard single site Metropolis-Hastings, for example), and then compute the posterior mean and variance of the above normal distribution. The last step is simply to draw sample from this distribution.

Next, we move on to inference for $m(\cdot)$ at unsampled inputs $x^*$ by putting together the classifier introduced in Section \ref{subsec:classification} and the response surface model discussed in this section.

\subsection{Combining local models}\label{subsec:local_emu}

We now put together the pieces of the fast local emulator for simulators with unknown constraints. We are interested in emulating $m(x^*)$, with simulation data $D = (\mbs{X},\mbs{m})$. The following steps are taken:

\begin{enumerate}[(i)]
    \item\label{lcGP:step1} Standardization:
    
    Following the convention adopted in GP modeling \citep{higdon2008computer}, the input region is mapped to the $d$-dimensional unit cube $[0,1]^d$.
    
    \item Local classification: Algorithm \ref{ag:y}
    
    A $n$-run nearest neighbor design for $x^*$ is constructed from $D$. If the outputs of this design share the same class label, $y(x^*)$ is predicted to be that label with probability 1. Otherwise, the model described in Section \ref{subsec:classification} is used to estimate $y(x^*)$. Denote by $\hat{f}(x^*)$ and $\hat{y}(x^*)$ the resulting emulator for $f(x^*)$ and $y(x^*)$, respectively.

    \item \label{lcGP:step3}Response surface model: Algorithm \ref{ag:z}
    
    When $\hat{y}(x^*)=1$, we move on to the local response surface model. Another set of local data consisting of active points, denoted by $A(x^*)$, is constructed as described in Algorithm \ref{ag:z}. With $A(x^*)$ in hand, the predictive distribution of $z(x^*)$ can be obtained as in \eqref{eq:Z_post}, with the prediction denoted $\hat{z}(x^*)$.
    
    \item Putting everything together:

    The resulting emulator for $m(x^*)$, denoted by $\hat{m}(x^*)$, has two components: $\hat{y}(x^*)$ and $\hat{z}(x^*)$:
    \begin{equation}\label{eq:map_m}
    \hat{m}(x^*) = \left\{\begin{array}{ll} \textrm{NA}\ \ \ \ \ \ \ \  & \textrm{  if } \hat{y}(x^*) = 0\\
    \hat{z}(x^*)\ \ \ \ \ \ \ \  & \textrm{  if } \hat{y}(x^*) = 1\end{array}\right..
    \end{equation}
\end{enumerate}

Up until now, we have been somewhat vague about how to predict $y(x^*)$ after sampling $\hat{f}(x^*)$ from its posterior distribution. In practice, several approaches can be taken to generate a prediction of $y(x^*)$. One way, for example, is to set $\hat{y}(x^*)=1$ if its posterior mean is larger than a user specified threshold (we use 0.5 in examples later) and zero otherwise. Alternatively, one could use the MAP (maximum a posteriori) estimate. A third way that adopts the MAP estimate of binary classification outcome is used later in Section \ref{sec:emu_example}.

In some applications, an unsuccessful simulation corresponds to zero instead of NA as output. To emulate $m(x^*)$ in this setting, samples of $\hat{f}(x^*)$ and $\hat{z}(x^*)$ are drawn as outlined after \eqref{eq:logodds_post} and \eqref{eq:Z_post}. A random sample of $\hat{m}(x^*)$ is set to the weighted average between zero and $\hat{z}(x^*)$, with probabilities $\frac{e^{\hat{f}(x^*)}}{1+e^{\hat{f}(x^*)}}$ and $\frac{1}{1+e^{\hat{f}(x^*)}}$:
\begin{equation}\label{eq:mean_m}
\hat{m}(x^*) = \frac{e^{\hat{f}(x^*)}}{1+e^{\hat{f}(x^*)}}\cdot \hat{z}(x^*) + \frac{1}{1+e^{\hat{f}(x^*)}}\cdot 0 = \frac{e^{\hat{f}(x^*)}}{1+e^{\hat{f}(x^*)}}\cdot \hat{z}(x^*).
\end{equation}

By adopting local models, the argument is that far away points contribute negligibly little to the prediction at $x^*$ relative to the neighboring points.  The local classifier proposed extends the nearest neighbor classification method \citep{cover1967nearest} with a GP classifier \citep{williams2006gaussian} to consider anisotropic constraints and to offer a local assessment of uncertainty. The proposed response surface model attempts to address local anisotropic behavior in $z(\cdot)$, and it is also possible to incorporate more sophisticated selection criteria \citep{gramacy2015local} if enough active local points are available.

An important aspect of the proposed method is that it is highly parallelizable. The emulation of individual points of interest $x^*$ can be easily extended to a large number of points by employing multiple CPUs and distributing the points among the CPUs. If needed, the classification and response surface model step can also be parallelized in light of the independence assumption for $y(x)$ and $z(x)$. 
Another convenient feature of the model is that it allows us to tackle the sequential design problem for computer models with unknown constraints. In the next section, a design criterion is introduced to guide future simulations.

\subsection{Sequential design}\label{subsec:sequential}
A practical problem of interest is to select new simulation runs to improve the emulator. In this section, we propose a design criterion, conventionally called an improvement function, that aims to reduce misclassification in $\hat{y}(\cdot)$ and also the predictive variance of $\hat{z}(\cdot)$ for computer models with unknown constraints. Let $\tilde{x}$ be a candidate design point. Improvement functions, denoted by $I(\tilde{x})$, have been used for {\it sequential design} to achieve various goals such as optimization \citep{jones1998efficient} and contour estimation  \citep{bingham2014design}.  

Since $m(\tilde{x})$ is labeled as `NA' when $y(\tilde{x})=0$, the variability of $\hat{m}(\tilde{x})$ can not be derived directly. Here, we redefine the predictive variance of $\hat{m}(\tilde{x})$, $\textrm{Var}[\hat{m}(\tilde{x})]$, as the generalized predictive variance of $(\hat{y}(\tilde{x}),\hat{z}(\tilde{x}))'$, written as
\begin{equation}\textrm{Var}[\hat{m}(\tilde{x})] = \left | \textrm{Var}\left[\begin{pmatrix} \hat{y}(\tilde{x})\\\hat{z}(\tilde{x})\end{pmatrix}\right] \right | = \left|\begin{matrix}\textrm{Var}[\hat{y}(\tilde{x})] & 0 \\ 0 & \textrm{Var}[\hat{z}(\tilde{x})]\end{matrix}\right| = \textrm{Var}[\hat{y}(\tilde{x})]\cdot \textrm{Var}[\hat{z}(\tilde{x})]\label{eq:var_general}\end{equation}
An intuitive sequential design procedure is to minimize the maximum predictive variance by assigning new simulations where $\textrm{Var}(\hat{m}(\tilde{x}))$ is the largest. Note that $\textrm{Var}(\hat{y}(\tilde{x}))$ is largest when $\textrm{P}(\hat{y}(\tilde{x})=1)=0.5$, which means that any sequential strategy that attempts to reduce $\textrm{Var}(\hat{y}(\tilde{x}))$ will tend to place new simulations near the constraint boundary. On the other hand, $\textrm{Var}(\hat{z}(\tilde{x}))$ is larger in the case of extrapolation, meaning that $\tilde{x}$ with larger $\textrm{Var}(\hat{z}(\tilde{x}))$ resides 
far from the center of the local design. Therefore, larger $\textrm{Var}(\hat{m}(\tilde{x}))$ corresponds to $\tilde{x}$ that is near the boundary of $C$. The end result is overemphasizing the improvement of $\hat{y}(\tilde{x})$, and placing little emphasis on improving predictions of $\hat{z}(\tilde{x})$.

Alternatively, for computer models that produce zero as an output when $x\not\in C$, the computer model can be written as $m(x)=y(x)\cdot z(x)$. Here, we standardize the simulation output to the unit interval to prevent the variance of $\hat{m}(\tilde{x})$ from being dominated by the scale of $\hat{z}(\tilde{x})$, and to put $\hat{y}(\tilde{x})$ and $\hat{z}(\tilde{x})$ on a similar scale. The predictive variance of $\hat{m}(\tilde{x})=\hat{y}(\tilde{x})\cdot \hat{z}(\tilde{x})$ can be written as
\begin{eqnarray}
\textrm{Var}(\hat{m}(\tilde{x})) & = & \textrm{E}(\textrm{Var}(\hat{m}(\tilde{x})|\hat{y}(\tilde{x}))) + \textrm{Var}(\textrm{E}(\hat{m}(\tilde{x})|\hat{y}(\tilde{x}))) \nonumber\\
& = & \textrm{E}(\hat{y}^2(\tilde{x}))\cdot \textrm{Var}(\hat{z}(\tilde{x})) + \textrm{Var}(\hat{y}(\tilde{x}))\cdot \textrm{E}^2(\hat{z}(\tilde{x})) \nonumber\\
& = & \underbrace{\textrm{E}(\hat{y}(\tilde{x}))\cdot \textrm{Var}(\hat{z}(\tilde{x}))}_{\textrm{variation of response surface}} + \underbrace{\textrm{Var}(\hat{y}(\tilde{x}))\cdot \textrm{E}^2(\hat{z}(\tilde{x}))}_{\textrm{classification variation}}.
\label{eq:var_decomp}
\end{eqnarray}
By conditioning on $\hat{y}(\tilde{x})$, the predictive variance of $\hat{m}(\tilde{x})$ contains two components, either carrying the predictive variance of the classifier $\hat{y}(\tilde{x})$ or that of the emulator $\hat{z}(\tilde{x})$. Both $ \textrm{Var}(\hat{z}(\tilde{x}))$ and $\textrm{E}^2(\hat{z}(\tilde{x}))$ are always between 0 and 1. In our experience, $\textrm{Var}(\hat{y}(\tilde{x}))$ quickly becomes the dominant term in the sum in Equation \eqref{eq:var_decomp} for $\tilde{x}$ with large $\textrm{Var}(\hat{m}(\tilde{x}))$ as more simulations are performed. This leads to the majority of new simulations being placed near the constraint boundaries with relatively little consideration for improving predictions of ${z}(\tilde{x})$. We omit the demonstration of this phenomenon, but we have found that improvement functions that address only the total variance of $\hat{m}(\tilde{x})$ will produce less favorable results for the variance of the emulator $\hat{z}(\tilde{x})$. 

We propose the following improvement function, which can be viewed as an adaptation of the first term to a type of contour estimation; the goal is to focus on improving the estimate of $z$, but only if there is some reasonable probability that {$y(\tilde{x})$} is non-zero, i.e., that $\tilde{x}$ satisfies the constraint:
\begin{eqnarray}\label{eq:contour_ctn}
	I(\tilde{x}) & = &  \max(0,\gamma(\tilde{x}) + \hat{q}(\tilde{x}) - p_{\textrm{thres}})\cdot \textrm{Var}(\hat{z}(\tilde{x})),
\end{eqnarray}
where $\hat{q}(\tilde{x})=\frac{e^{\hat{f}(\tilde{x})}}{1 + e^{\hat{f}(\tilde{x})}}$, is an estimate of the probability of successful outcome at input $\tilde{x}$ given the simulation data, and $p_\textrm{thres}$ is a user-specific classification threshold. The difference between a percentile (we choose the 95th for example) and the mean of $\hat{q}(\tilde{x})$ is denoted $\gamma(\tilde{x})$. That is, $\textrm{P}(\hat{q}(\tilde{x})\leq \gamma(\tilde{x})+\textrm{E}[\hat{q}(
\tilde{x})]) = 0.95$. The second term $\textrm{Var}(\hat{z}(\tilde{x}))$ is the predictive variance of $\hat{z}(\tilde{x})$ given simulation data $D$, and reducing $\textrm{Var}(\hat{z}(\tilde{x}))$ will lead to better performance of $\hat{z}(\cdot)$. Recall from Sections \ref{subsec:classification} and  \ref{subsec:emulation} that the proposed method produces the posterior samples of $\hat{q}(\tilde{x})$ and $\textrm{Var}(\hat{z}(\tilde{x}))$, which allows us to obtain $\textrm{E}(I(\tilde{x}))$ directly. For candidate simulation points $(\tilde{x}_1,\tilde{x}_2,\ldots,\tilde{x}_K)$, one can then choose the trial with the largest $\textrm{E}(I(\tilde{x}_i))$ to perform future simulations. This improvement function combines the needs to detect unknown constraints and to explore active regions at the same time. The user-specified threshold $p_\textrm{thres}$ can be viewed as a tuning constant that trades-off having more predicted successful simulations (BBH mergers) with reducing output uncertainty within the constraint regions (the chirp mass of BBH mergers). A detailed illustration will be given later in Section \ref{subsec:2D_toy}.

\section{Numerical Illustrations }\label{sec:emu_example}
In this section, two synthetic examples are used to illustrate the proposed emulation method, that we call lcGP for {\em local constrained GP}. To compare with existing emulation methods, we use models where the output $m(x) \in \mathbb{R}^+ $ for $x\in C$, and $m(x)=0$ otherwise. The performance of the proposed approach is compared with the traditional GP, laGP  \citep{gramacy2015local}, and a global GP classifier coupled with a global GP. Section \ref{subsec:1D} provides a simple and intuitive example of a {\it top hat} function \citep{dunlop2018deep} to demonstrate the benefits of coupling the classification model $y(x)$ with the traditional GP $z(x)$. The second example (Section \ref{subsec:2D_toy}) illustrates in detail how to implement the proposed methodology for emulation along with sequential design.

Denote by $\mbs{X}^*$ the set of inputs to emulate. For each $x_i^*\in\mbs{X}^*$, Algorithm \ref{ag:y} is adopted to find $n$ neighboring points to $x_i^*$ in $D$, and to predict $y(x_i^*)$. If $\hat{y}(x_i^*)=0$, we predict $\hat{m}(x_i^*) = 0$. Otherwise, a local active design as described in Algorithm \ref{ag:z} is constructed, and local emulator $z(\cdot)$ is used to predict $\hat{m}(x_i^*) = \hat{z}(x_i^*)$. For different $x_i^*$'s, emulation can be conducted in parallel since the local models are independent. If fast emulation is the goal, one need only estimates (e.g., using the MAP estimate) for $\hat{y}(x_i^*)$ and $\hat{z}(x_i^*)$, respectively, to obtain $\hat{m}(x_i^*)=\hat{y}(x_i^*)\cdot \hat{z}(x_i^*)$. On the other hand, if uncertainty quantification of the computer model is of interest (as in the sequential design problem for example), posterior 
sampling of $\hat{f}(x^*_i)$ (Section \ref{subsec:classification}) and $\hat{z}(x_i^*)$ (Section \ref{subsec:emulation}) should be done. To provide a detailed illustration, the latter approach is taken for the two examples.

\subsection{The top hat function}\label{subsec:1D}
\begin{figure}[!h]
\captionsetup[subfigure]{font=small}
\begin{subfigure}{.3\linewidth}
\includegraphics[scale=.3]{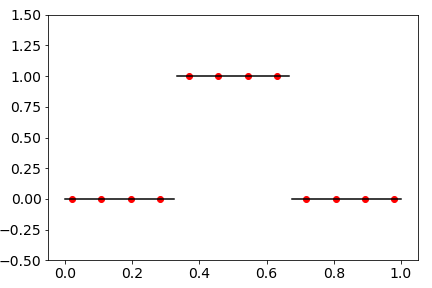}
\caption{The top hat function}\label{fig:tophat_true}
\end{subfigure}
\hspace{6pt}
\begin{subfigure}{.3\linewidth}
\includegraphics[scale=.3]{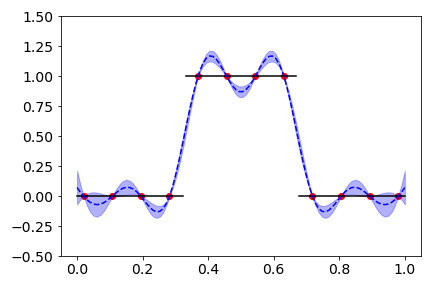}
\caption{GP}\label{fig:tophat_GP}
\end{subfigure}
\hspace{6pt}
\begin{subfigure}{.3\linewidth}
\includegraphics[scale=.3]{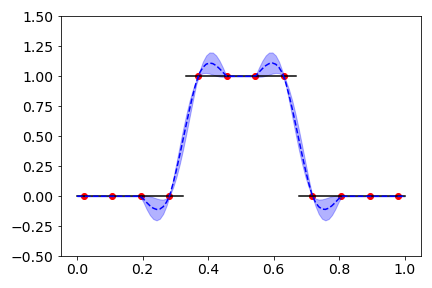}
\caption{laGP}\label{fig:tophat_laGP}
\end{subfigure}
\begin{subfigure}{.3\linewidth}
\includegraphics[scale=.3]{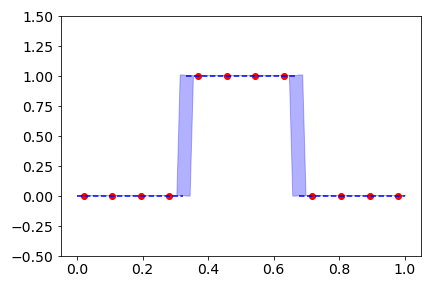}
\caption{lcGP (binary)}\label{fig:tophat_lcGP_map}
\end{subfigure}
\hspace{6pt}
\begin{subfigure}{.3\linewidth}
\includegraphics[scale=.3]{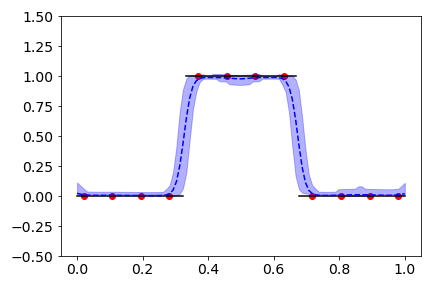}
\caption{lcGP (expected)}\label{fig:tophat_lcGP_mean}
\end{subfigure}
\caption{Emulation of the top hat function with different methods. Black lines represent the true function and red dots are the observed simulation data. Point-wise predictive intervals with 95\% confidence are highlighted with light blue color and dashed blue curves depict the predicted response. The predictions adopt the predictive means for the traditional GP method and the laGP method, and combine the maximum a posteriori (MAP) estimates for $\hat{y}(\cdot)$ and $\hat{z}(\cdot)$ for the lcGP method.}\label{fig:1d_examples}
\end{figure}

To provide intuition for the performance of the proposed method, we start by emulating the simple function \citep{dunlop2018deep} shown in Figure \ref{fig:1d_examples}. The true function (black lines) and simulation responses (red dots) are shown in Figure \ref{fig:tophat_true}. Two different predictive approaches for lcGP are considered. The first, referred to as the binary classification approach from here on, sets $\hat{y}(x^*)$ to one when $\hat{q}(x^*) > 0.5$ (Figure \ref{fig:tophat_lcGP_map}), and zero otherwise. The other sets $\hat{y}(x^*)$ equal to $\hat{q}(x^*)$, which is the expected value of $\hat{y}(x^*)$ given $\hat{q}(x^*)$ (Figure \ref{fig:tophat_lcGP_mean}).

For each of the emulation methods, we predict the model output at 100 evenly spaced points on the unit interval. {Our goal is to make predictions with associated predictive intervals based on the 12-run simulation shown as red dots in Figure \ref{fig:tophat_true}.} For laGP, a local design of size 4 is used. For the proposed lcGP method, emulation with both binary classification for {$\hat{y}(x^*)$} as in \eqref{eq:map_m} (Figure \ref{fig:tophat_lcGP_map}) and the expected value of {$\hat{y}(x^*)$ ($x^*\in\mbs{X}^*$)} as in \eqref{eq:mean_m} (Figure \ref{fig:tophat_lcGP_mean}) are produced. {A uniform prior distribution, $U(0,\sqrt{10})$, is chosen for the GP classifier length-scale parameter, $\phi$, and this corresponds to a maximum correlation of $0.9048$ between log-odds at the two extremes $x=0$ and $x=1$. The prior distribution for the inverse of the precision parameter, $\eta^{-1}$, is chosen as $U(0,4)$. With $\eta^{-1}=4$, the three-standard-deviation limits, $(-6,6)$, for the log-odds are equivalent to the log-odds of $(0.0025, 0.9975)$ for the classification probabilities. This suggests that $U(0,4)$ is a relatively uninformative prior distribution for $\eta^{-1}$. Elliptical slice sampling is used to sample the latent log-odds, and single site Metropolis-Hastings is used to sample the GP parameters, $\phi$ and $\eta^{-1}$. Python code used to for this example can be found in the supplemental materials \citep{supplement}.} To construct prediction intervals with lcGP, posterior samples of {$\hat{m}(x^*)$} are drawn by
\begin{enumerate}[(i)]
    \item sampling the classification probability $\hat{q}(x^*) = (1+\exp(-\hat{f}(x^*)))^{-1}$ as described in Section \ref{subsec:classification};
    \item generating posterior samples of $\hat{y}(x^*)$ ($0$'s and $1$'s) based on $\hat{q}(x^*)$ (Figure \ref{fig:tophat_lcGP_map}), or, setting $\hat{y}(x^*)$ to $\hat{q}(x^*)$ (Figure \ref{fig:tophat_lcGP_mean});
    \item drawing posterior samples of $\hat{z}(x^*)$ as described in Section \ref{subsec:emulation};
    \item obtaining the posterior samples for $\hat{m}(x^*) = \hat{y}(x^*)\cdot \hat{z}(x^*)$.
\end{enumerate}

Pointwise $95\%$ prediction intervals are constructed directly with the resulting posterior samples of $\hat{m}(x^*)$. Figure \ref{fig:1d_examples} illustrates that the traditional GP and laGP attempt to smooth out the discontinuity, thereby causing the emulator to perform relatively poorly on the constant regions. The proposed lcGP methods are able to more closely emulate the true behavior of the computer model within the constraint region, and identify the uncertainty near the constraint boundaries. {It is worth-noting that the uncertainty captured by the confidence interval of the lcGP method (Figures \ref{fig:tophat_lcGP_map} and \ref{fig:tophat_lcGP_mean}) does not include that of the Bernoulli distribution of $\hat{y}(x^*)$. Rather, it addresses the posterior uncertainty of the latent log-odds variable $\hat{f}(x^*)$.}

\subsection{A Two-dimensional Example}\label{subsec:2D_toy}
\begin{figure}[!t]
    \centering
    \begin{subfigure}[t]{0.55\linewidth}
    \centering
    \includegraphics[scale=.5]{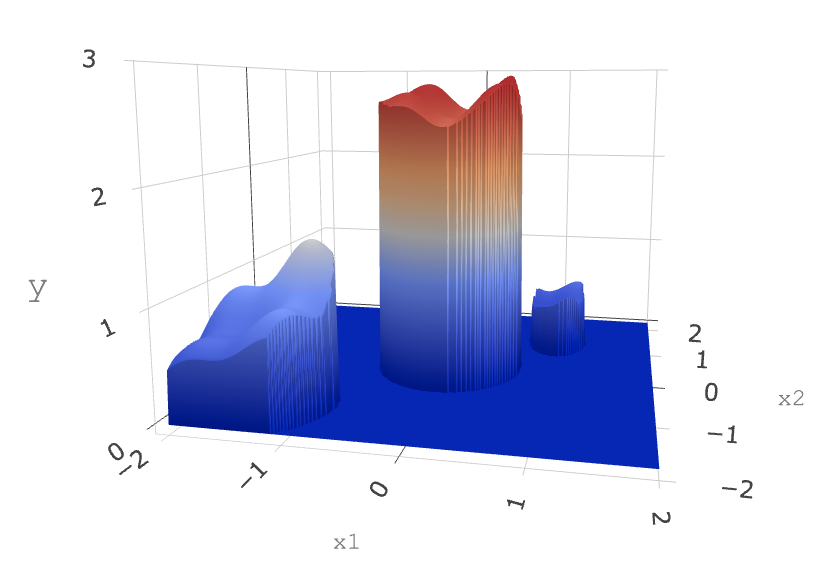}
    \caption{True function output with constraints}\label{fig:true_fun}
    \end{subfigure}
    \begin{subfigure}[t]{0.4\linewidth}
    \centering
    \includegraphics[scale=.38]{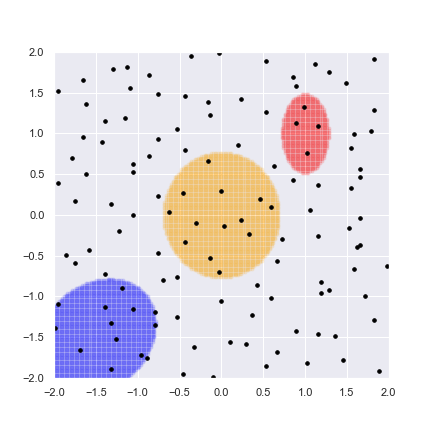}
    \caption{Input space and design}\label{fig:design144}
    \end{subfigure}
    \caption{A 2D example with unknown constraints}
\end{figure}

Consider an example model from \citet{gramacy2015local} that was presented without constraints:
\begin{align*}
    \eta(x_1,x_2) = & \big(e^{-(x_1-1)^2} + e^{-0.8(x_1+1)^2}-0.1\sin(8(x_1+0.1))\big) \\
    & \cdot \big(e^{-(x_2-1)^2} + e^{-0.8(x_2+1)^2}-0.1\sin(8(x_2+0.1))\big),
\end{align*}
where $(x_1,x_2)\in [-2,2]^2$. Three constraint regions (highlighted areas in Figure \ref{fig:design144}) are  imposed on $\eta(\cdot)$ so that the input space has three regions with positive output and the rest of the input space results in zero output. We also elevate or lower the response surface for different regions by three constants to generate greater model variability amongst the regions (see Figure \ref{fig:true_fun}). The resulting computer model is:

\begin{equation}
     m(x_1,x_2) = \left\{
  \begin{array}{ll}
  -0.5 + \eta(x_1,x_2)  & \textrm{ when } 10(x_1-1)^2 + (x_2-1)^2/0.3 < 0.8 \textrm{ (the red region),}\\
   0.2 + \eta(x_1,x_2)  & \textrm{ when }(x_1+x_2+3)^2/1.6 + (x_1-x_2)^2 < 0.8 \textrm{ (the blue region),}\\
   1.8 + \eta(x_1,x_2)  & \textrm{ when }x_1^2/0.9 + x_2^2 < 0.6 \textrm{ (the yellow region),}\\
  0  & \textrm{ otherwise.}\\
  \end{array}
    \right.
\end{equation}

A randomly generated 121-run orthogonal array based Latin hypercube design \citep{tang1993orthogonal} (black dots in Figure \ref{fig:design144}) is chosen for simulation inputs, and a $71\times 71$ grid
is placed on the input domain $[-2,2]^2$ to construct a validation set to assess the performance of different emulation approaches. For lcGP, the local design sizes are $n=n^M=12$.

\begin{figure}[!t]
\centering
\begin{subfigure}[t]{.32\linewidth}
\includegraphics[width=1.0\linewidth]{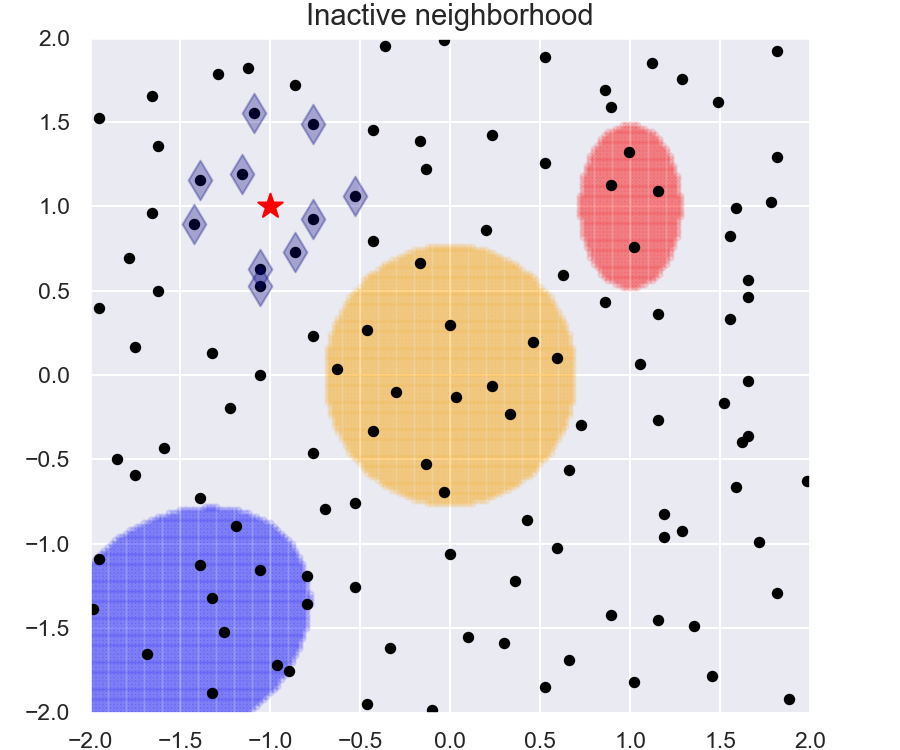}
\caption{Inactive neighborhood}\label{fig:local_inact}
\end{subfigure}
\begin{subfigure}[t]{.32\linewidth}
\includegraphics[width=1.0\linewidth]{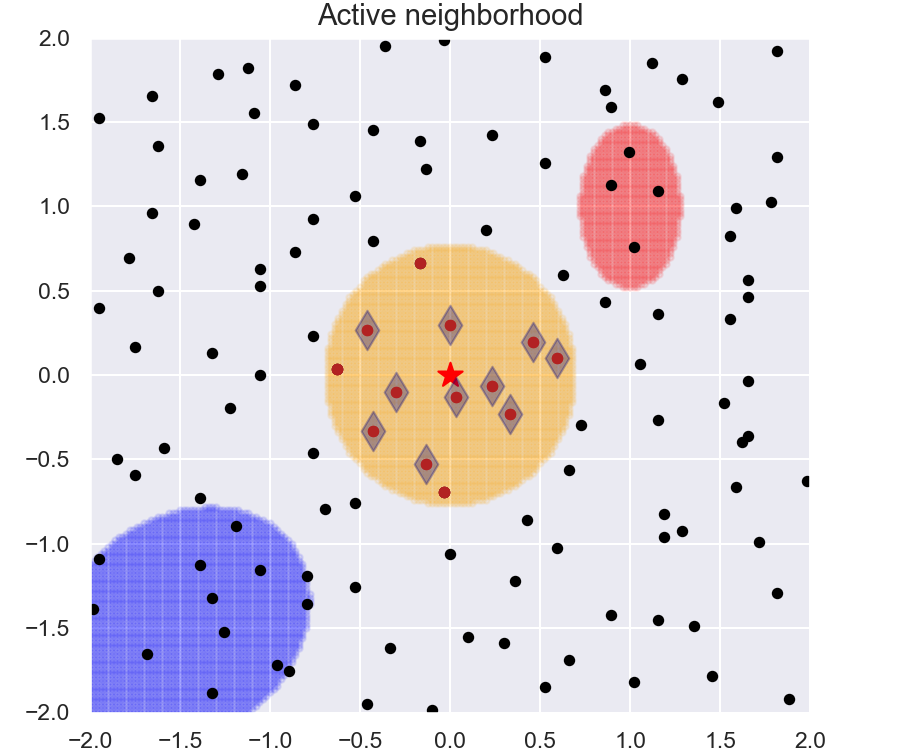}
\caption{Active neighborhood}\label{fig:local_act}
\end{subfigure}
\begin{subfigure}[t]{.32\linewidth}
\includegraphics[width=1.0\linewidth]{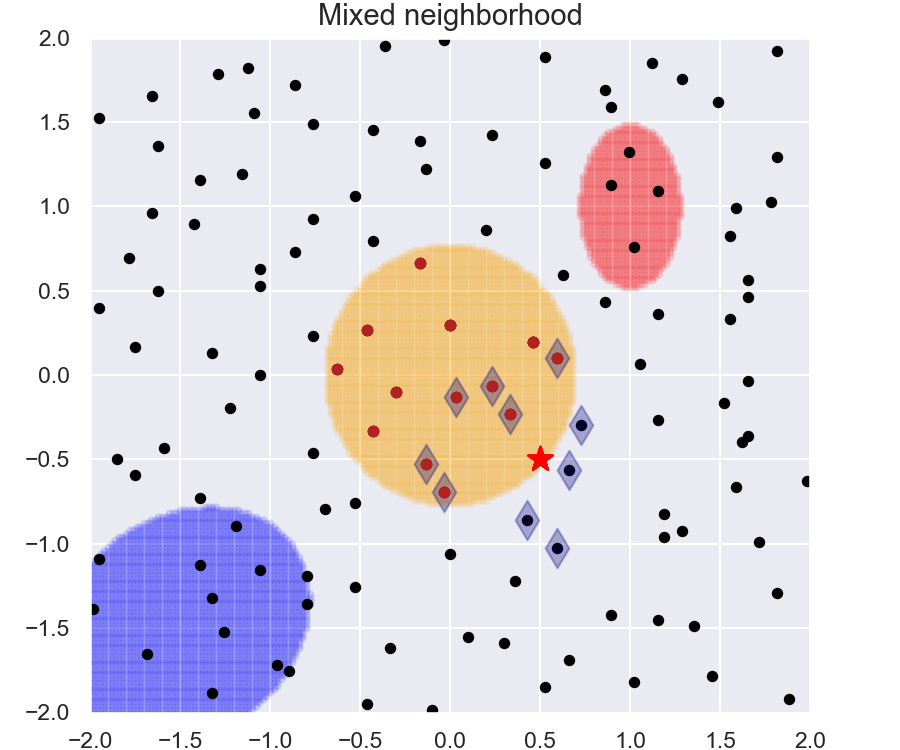}
\caption{Mixed neighborhood}\label{fig:local_mixed}
\end{subfigure}
\caption{Local designs for different inputs. The red stars represent the points to emulate, blue diamonds are local designs used for classification and red dots are local designs used for response surface model.}\label{fig:local_design}
\end{figure}

Figure \ref{fig:local_design} shows the local designs constructed for different inputs. To get an idea of how the local designs are selected in practice, there are three scenarios considered for emulation at an input $x^*$: 
\begin{enumerate}[(i)]
\item when $x^*$ (red star) is surrounded entirely by design points (blue diamonds) that result in a zero output (Figure \ref{fig:local_inact}), the classifier will predict $\hat{y}(x^*)=0$ and $\hat{m}(x^*)=0$ with probability one;
\item when all neighboring design points of $x^*$ are active (Figure \ref{fig:local_act}),  $\hat{q}(x^*)=1$ and $z(x^*)$ is emulated with an expanded active design (red dots).
\item when $x^*$ is in a mixed neighborhood (Figure \ref{fig:local_mixed}), the classification of $y(x^*)$ with the local design (blue diamonds) is followed by the emulation of $z(x^*)$ with the expanded active design (red dots).
\label{sci:mixed}
\end{enumerate}

{Similar to the previous example, a uniform prior distribution, $U(0,\sqrt{10})$, is chosen for the length-scale parameter, $\phi$, of the local GP classifier, and a $U(0,4)$ is used for the prior distribution on $\eta^{-1}$. Elliptical slice sampling is used for the latent log-odds, and single site Metropolis-Hastings is used to sample the GP parameters. We found $3000$ MCMC iterations, with $1000$ steps serving as  burn-in, to be successful for this example. The average time to emulate each $m(x^*)$ is roughly 1.2 seconds on an Intel Core i7 processor and 16GB of memory. Implementation details and MCMC diagnostics can be found in the supplemental materials \citep{supplement}.} 

\begin{figure}[!t]
\centering
\begin{subfigure}{.24\linewidth}
\includegraphics[scale=.25]{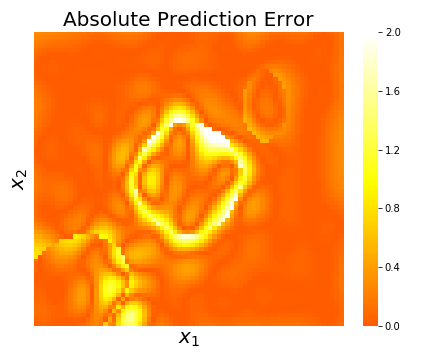}
\caption{Traditional GP}\label{fig:err_GP}
\end{subfigure}
\hspace{6pt}
\begin{subfigure}{.24\linewidth}
\includegraphics[scale=.25]{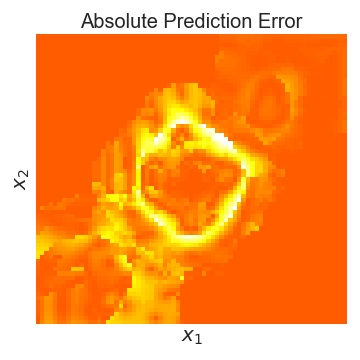}
\caption{laGP}\label{fig:err_laGP}
\end{subfigure}
\begin{subfigure}{.24\linewidth}
\includegraphics[scale=.25]{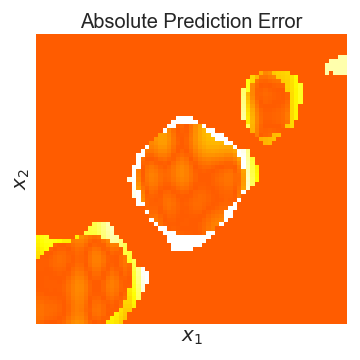}
\caption{Global GP\&GPC}\label{fig:err_lcGP_bayes}
\end{subfigure}
\begin{subfigure}{.24\linewidth}
\includegraphics[scale=.25]{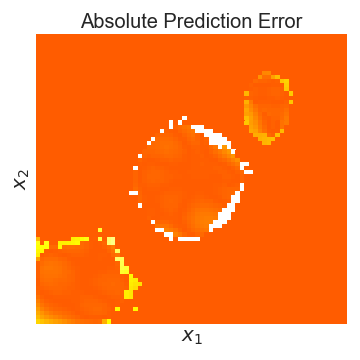}
\caption{lcGP}\label{fig:err_lcGP_trunc}
\end{subfigure}

\begin{subfigure}{.24\linewidth}
\includegraphics[scale=.25]{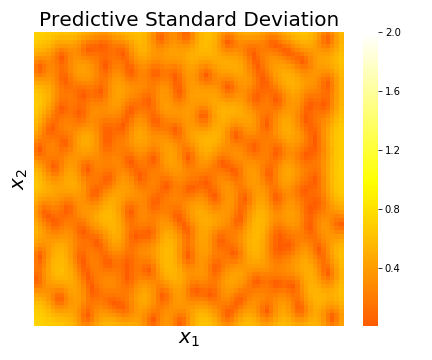}
\caption{Traditional GP}\label{fig:psd_GP}
\end{subfigure}
\hspace{6pt}
\begin{subfigure}{.24\linewidth}
\includegraphics[scale=.25]{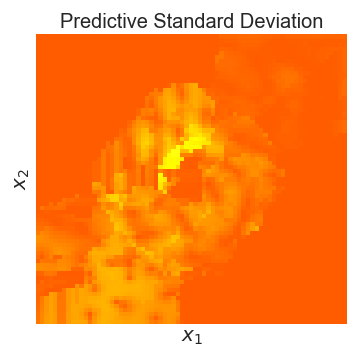}
\caption{laGP}\label{fig:psd_laGP}
\end{subfigure}
\begin{subfigure}{.24\linewidth}
\centering
\includegraphics[scale=.25]{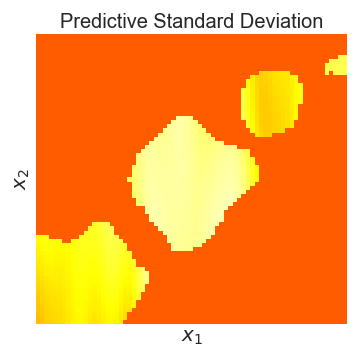}
\caption{Global GP\&GPC}\label{fig:psd_lcGP_bayes}
\end{subfigure}
\begin{subfigure}{.24\linewidth}
\includegraphics[scale=.25]{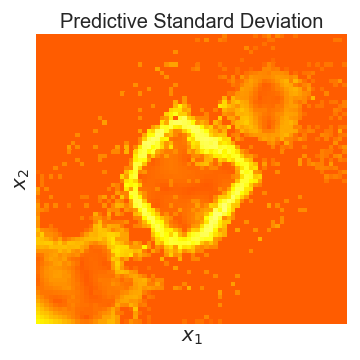}
\caption{lcGP}\label{fig:psd_lcGP_trunc}
\end{subfigure}
\caption{Heatmap of absolute prediction error (first row) and predictive standard deviation (second row) for different models trained with data collected from a 121-run orthogonal array based Latin hypercube design. Brighter colors indicate larger errors and greater predictive standard deviations. Figures \ref{fig:err_GP} and \ref{fig:psd_GP} are produced with the traditional GP model; Figures \ref{fig:err_laGP} and \ref{fig:psd_laGP} are produced with local approximate GP (laGP) method; Figures \ref{fig:err_lcGP_trunc} and \ref{fig:psd_lcGP_trunc} are produced with the proposed lcGP method; Figures \ref{fig:err_lcGP_bayes} and \ref{fig:psd_lcGP_bayes} adopt a model that combines a global GP classifier with a global GP model. Each row shares the same color bar as given in the plots of the first (left) column.}\label{fig:err_comp}
\end{figure}

In this example, the binary classification implementation of lcGP is used. The prediction for $\hat{y}(x^*)$ is chosen to be the MAP estimate based on posterior samples of $\hat{y}(x^*)$. Absolute prediction errors and predictive standard deviations are compared in Figure \ref{fig:err_comp} among different methods: the traditional GP (first column), laGP (second column), the combination of a global GP classifier and a global GP (third column) and lcGP (last column). In the heatmaps brighter/hotter color corresponds to larger absolute prediction error and larger predictive standard deviation. In the presence of unknown constraints, ignoring the constraints as in the traditional GP method (Figure \ref{fig:err_GP} and \ref{fig:psd_GP}) results in a considerable level of error and predictive uncertainty even at input locations that are far from the constraint boundaries. This phenomenon is likely caused by the fact that GP models attempt to explain the local sudden changes between zero and positive output with a small length-scale parameter and large GP variance. The laGP model (Figures \ref{fig:err_laGP} and \ref{fig:psd_laGP}) can help reduce the errors at locations that are far enough from the constraint boundaries, but it faces the same challenge as the traditional GP method for input locations that have both zeros and positive outputs in their local designs. For the proposed methods that combine a classifier and an emulator (Figures \ref{fig:err_lcGP_trunc},\ref{fig:err_lcGP_bayes}, \ref{fig:psd_lcGP_trunc} and \ref{fig:psd_lcGP_bayes}), it is evident that the local model (Figures \ref{fig:err_lcGP_trunc} and \ref{fig:psd_lcGP_trunc} ) outperforms the global one (Figures \ref{fig:err_lcGP_bayes} and \ref{fig:psd_lcGP_bayes}) by providing higher prediction accuracy and lower predictive uncertainty.

\begin{table}[!ht]
    \caption{Prediction accuracy for different methods}
    \label{tab:err_by_method_2D}
    \centering
    \begin{tabular}{lcccccc}
    \hline
      & \multicolumn{2}{c}{Misclassification} & NSE & NSE & RMSE & MAE \\
    \cmidrule{2-3}
    Method & active &inactive & active & all & active & active \\
     \hline
    traditional GP & na & na & $91.45\%$ & $78.92\%$ & $0.4601$ & $2.0316$ \\
    laGP &  na & na & $93.80\%$ & $81.38\%$ & $0.3915$ & $1.8211$ \\
    Global GP\&GPC & $10.91\%$ & $3.65\%$ & $97.28\%$ & $68.42\%$ &$0.1486$ & $0.8222$\\
    lcGP & $8.77\%$ & $1.86\%$ & $98.35\%$ & $81.44\%$& $0.1167$ & $0.8170$ \\
    \hline
    \end{tabular}
\end{table}

To further compare the methods, performance metrics including the Nash–Sutcliffe efficiency (NSE), the root-mean-square error (RMSE) and the maximum absolute error (MAE)
are computed and displayed in Table \ref{tab:err_by_method_2D}. Similar to the coefficient of determination, the NSE \citep{NASH1970282} attempts to measure the proportion of variation that can be explained by a predictive model. Here, NSEs for correctly classified active points and all data (correctly classified or not) are reported separately (columns 'NSE (active)' and 'NSE (all)') to examine the performance of $\hat{z}(x^*)$ and $\hat{m}(x^*)$ separately. For RMSE and MAE, results are reported on $\hat{z}(x^*)$.

It can be seen that the lcGP model outperforms the other three models across most predictive accuracy measurements. In this simple example with a small data set, the global model (classifier and emulator) does perform comparably well. However in problems with large data, the global model is often unattainable due to the computational burden discussed earlier.

\subsubsection*{Sequential Design}

\begin{figure}
    \centering
    \includegraphics[scale=.35]{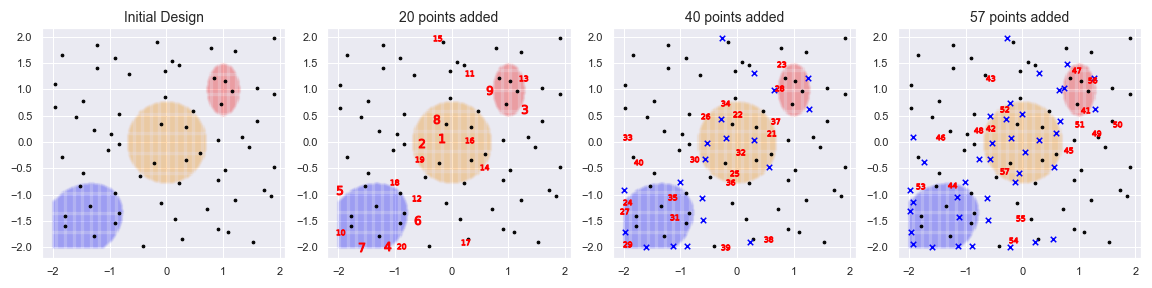}
    \caption{Sequential design construction, every 20 iterations. Black dots represent the initial design, blue x's are previously added simulations and red numbers represent the order in which new simulations are added.}
    \label{fig:seq_2D}
\end{figure}

Finally, the two-dimensional example is used to demonstrate how additional trials may be added to improve the model based on the improvement function introduced in Section \ref{subsec:sequential}. A 64-run orthogonal array based Latin hypercube design (first panel in Figure \ref{fig:seq_2D}) is used to conduct an initial set of simulations. An lcGP model is then constructed based on this initial design and corresponding outputs. To sequentially add new simulation runs, the following steps are taken at iteration $k$, $k=1,2,\ldots,57$:
\begin{enumerate}[(i)]
\item divide the input space $[-2,2]^2$ into a $20\times20$ grid and draw one random sample within each grid cell to form a candidate set of size 400, $(\tilde{x}_1,\tilde{x}_2,\ldots,\tilde{x}_{400})$, for new simulations.
\item calculate $\textrm{E}(I(\tilde{x}_i))$ for each candidate point and find the best candidate $\tilde{x}_k$ to perform the next simulation; obtain $\tilde{m}_k = m(\tilde{x}_k)$.
\item update the lcGP model with the newly simulated data point  $(\tilde{x}_k,\tilde{m}_k)$ added.
\end{enumerate}

To compare the sequential design with the 121-run design in Figure \ref{fig:local_design}, the search for new design points continues until a total run size of 121 is obtained. Figure \ref{fig:seq_2D} shows the updated design for every 20 iterations, with the final design plotted in the fourth panel. New points are labeled (in red) according to the order in which they are added. Previously added points are labeled by blue x's.  The sequential criterion mostly chooses points near the boundary of the constraint regions, while also jumping to sparsely sampled regions (e.g., points 15 and 33), and points within the constraint region (e.g., points 1 and 32) in an explore-and-exploit manner.

Next, we emulate, with lcGP, the computer model on the same $71\times71$ grid, with simulation data from the sequential design. The prediction accuracy is compared in Table \ref{tab:err_by_design_2D}  to that of the orthogonal array based Latin hypercube design with the same simulation size (Figure \ref{fig:local_design} and Table \ref{tab:err_by_method_2D}). By running more simulations on active regions, the sequential design is able to further reduce the RMSE for active points while maintaining similar classification accuracy.

\begin{table}[!ht]
    \caption{Prediction accuracy for different designs}
    \label{tab:err_by_design_2D}
    \centering
    \begin{tabular}{lcccccc}
    \hline
      & \multicolumn{2}{c}{Misclassification} & NSE & NSE & RMSE & MAE \\
    \cmidrule{2-3}
    Method & active &inactive & active & all & active & active \\
     \hline
    Latin hypercube design & $8.77\%$ & $1.86\%$ & $98.35\%$ & $81.44\%$& $0.1167$ & $0.8170$ \\
    Sequential design $E(I(x))$ & $9.05\%$ & $1.99\%$ & $99.71\%$ & $76.10\%$ & $0.0488$ & $0.4093$ \\
    \hline
    \end{tabular}
\end{table}

\section{Results for the COMPAS model}\label{sec:COMPAS_res}

We now return to the COMPAS model and apply the proposed emulation method to computer model runs produced using the STROOPWAFEL procedure\footnote{Available online at  \href{https://zenodo.org/record/3627403}{https://zenodo.org/record/3627403}. The simulation data used here is a combination of the exploration and refinement phase for BH-BH mergers.} \citep{broekgaarden2019stroopwafel}. In total, one million simulations of BBH formation were performed.  For these runs, the hyper-parameters $t$ of the COMPAS model were kept constant ($Z=0.001$, $\alpha=1$, $\sigma=265$km/s, $\textrm{flbv}=1.5$) and initial conditions $x$ were drawn using adaptive importance sampling.  These modifications made it possible to increase the BBH yield to approximately $27\%$, as samples are placed more densely in constraint-satisfying regions. In this example, the input dimension was reduced to 11 (the dimensionality of initial conditions $x$) since $t$ is fixed. For illustration of the emulation method proposed in Section \ref{subsec:COMPAS_emu}, we do not consider the distribution of $x$ until later (Section \ref{subsec:chirpMass_dist}) when we propagate the input distributions through the emulator in order to obtain the chirp mass distribution. 

\begin{figure}[h]
\begin{center}
\begin{subfigure}[t]{.47\linewidth}
\centering
\includegraphics[scale=.3]{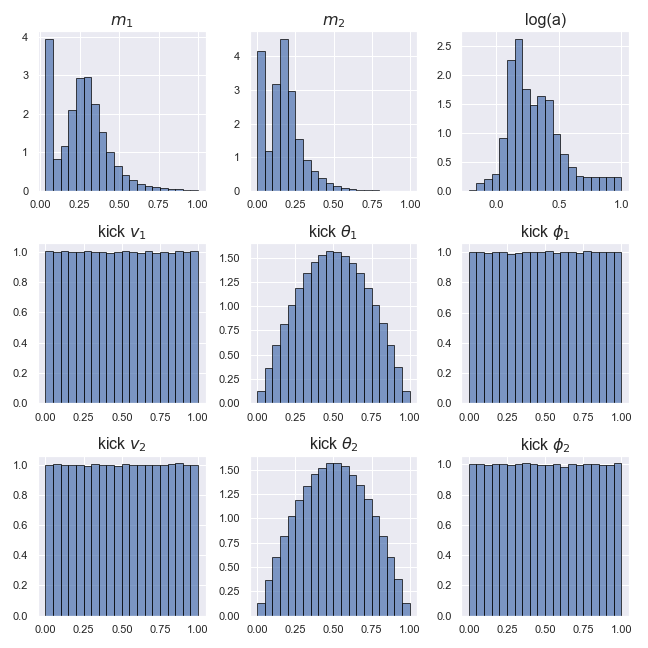}
\caption{Histogram of selected COMPAS design inputs from the STROOPWAFEL simulations after standardization.}\label{fig:input_hist}
\end{subfigure}
\hspace{16pt}
\begin{subfigure}[t]{.47\linewidth}
\centering
\includegraphics[scale=.3]{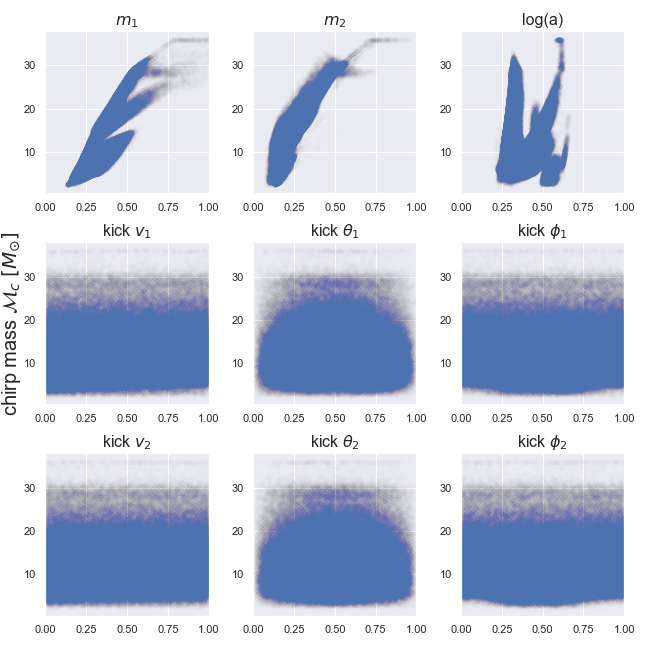}
\caption{Scatter plots of the chirp mass versus each input. Opaque regions correspond to high concentration of active simulations in the data, while transparent regions correspond to few active simulations. }\label{fig:input_mass}
\end{subfigure}
\end{center}
\caption{Visualization of COMPAS simulation data}\label{fig:COMPAS_summary}
\end{figure}

\begin{figure}
    \centering
    \includegraphics[scale=.235]{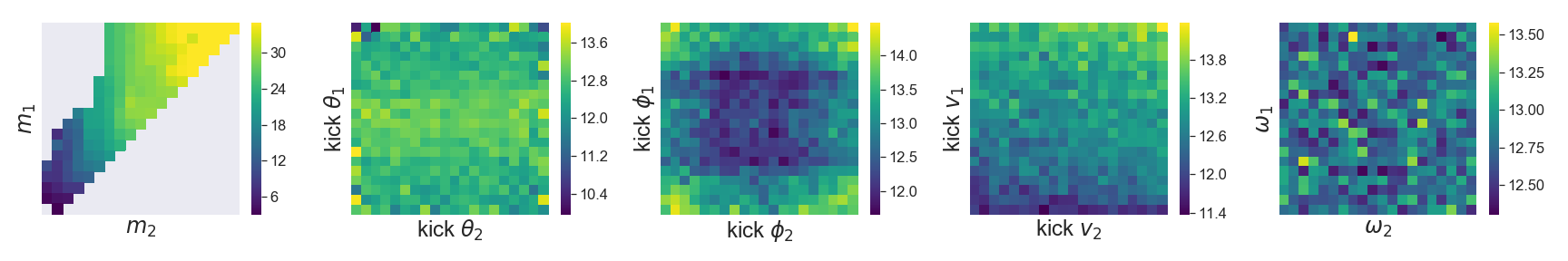} 
    \caption{Heatmaps of the chirp mass against pairs of inputs.}
    \label{fig:chirpMass_heatmap}
\end{figure}

Figure \ref{fig:COMPAS_summary} and Figure \ref{fig:chirpMass_heatmap} provide a    visualization of the simulation data standardized to the unit hypercube $[0,1]^{11}$. Histograms of selected standardized input variables from all simulations (Figure \ref{fig:input_hist}) and scatter plots of the BBH chirp mass against individual input variables (Figure \ref{fig:input_mass}) for successful simulations are shown. Heatmaps of the chirp mass against pairs of input parameters are shown in Figure \ref{fig:chirpMass_heatmap}. The histograms show that $x$ is non-uniform across the input space. The resulting space filling might be unsatisfactory and could bring considerable uncertainty into emulation. For the scatter plots, we observe some trends between the chirp mass and the initial component masses $m_1$, $m_2$, and separation, while the kick variables seem to have little impact on the value of the chirp mass. However, the kick parameters can impact the outcome of the model, i.e., whether a merging BBH is formed. Next, we illustrate in detail how emulation of the COMPAS data is performed and evaluated.

\subsection{Emulation}\label{subsec:COMPAS_emu}
For computer models like COMPAS, where the input variables differ largely in scale and physical meaning (Table \ref{tab:data}), a {\it{stretching and compressing}} procedure \citep{hsu2019fast} is added to step \ref{lcGP:step1} of the procedure described in Section \ref{subsec:local_emu}. This procedure aims to modify the definition of distance in the input space, based on the correlation between simulation outputs, so that the concept of `neighborhood' in lcGP can incorporate informative local points. More specifically for this example, 1000 active simulations are randomly selected and modeled using an anisotropic GP with squared-exponential covariance. This procedure is repeated 100 times, and the average estimated length-scale, $\mbs{\kappa}$, is used to scale $\mbs{x}$ by setting $\mbs{{x'}} = \mbs{x}/ \mbs{\kappa}$. The construction of local designs in the lcGP method (as in Algorithms \ref{ag:y} and \ref{ag:z}) is based on $\mbs{{x'}}$, while the emulator model itself takes the standardized input $\mbs{x}$. {For model input $\mbs{x}=(m_1,m_2,\log(a),v_1,v_2,\theta_1,\theta_2,\phi_1,\phi_2, \omega_1, \omega_2)$, the value for $\mbs{\kappa}$ is chosen as $(0.1,0.1,0.1,1,1,1,1,1,1,1,1)$. This choice of value for $\mbs{\kappa}$ implies that for chirp mass the spatial correlations with respect to the initial masses and separation are larger than those of other input dimensions.}

Since the chirp mass of a BBH is always positive, we introduce a left truncation to step \ref{lcGP:step3} in Section \ref{subsec:local_emu} for $\hat{z}(x^*)$ as
\begin{equation}
\label{eq:pred_trunc}
p(\hat{z}(x^*)|\hat{z}(x^*)>0,\mbs{z}^{ab}) = \frac{p(\hat{z}(x^*)|\mbs{z}^{ab})}{P(\hat{z}(x^*)>0|\mbs{z}^{ab})},
\end{equation}
where $p(\hat{z}(x^*)|\mbs{z}^{ab})$ is as in Equation \eqref{eq:Z_post}. We choose to apply the truncation after building $\hat{z}(\cdot)$ for speedy emulation. Similar to the examples in Section \ref{sec:emu_example}, the MAP estimate is used for $\hat{y}(x^*)$ and $\hat{z}(x^*)$.

To evaluate the performance of the method proposed, we performed $100$ independent cross validations with a holdout size of $1000$ drawn by simple random sampling. The local design size $n$ and $n^M$ are chosen to be $50$. To achieve fast emulation, we also estimate beforehand and fix the GP length-scale parameters for both the classification and the response surface model here.
  This procedure requires in total \num{100000} individual emulations which required $3.25$ hours when parallelized on $10$ nodes of Intel E5-2683 v4 "Broadwell" at 2.1Ghz with 8GB memory. Thus, one emulation required an average of only $\approx 0.1$s in this regime. {It is worth-noting that larger choice of $n$ has also been investigated. With $n$ is raised to $100$, the computation time triples on the same machine while the misclassification is similar. Therefore $n=50$ is considered appropriate for this demonstration.}

\begin{figure}[!b]
    \centering
\includegraphics[scale=.45]{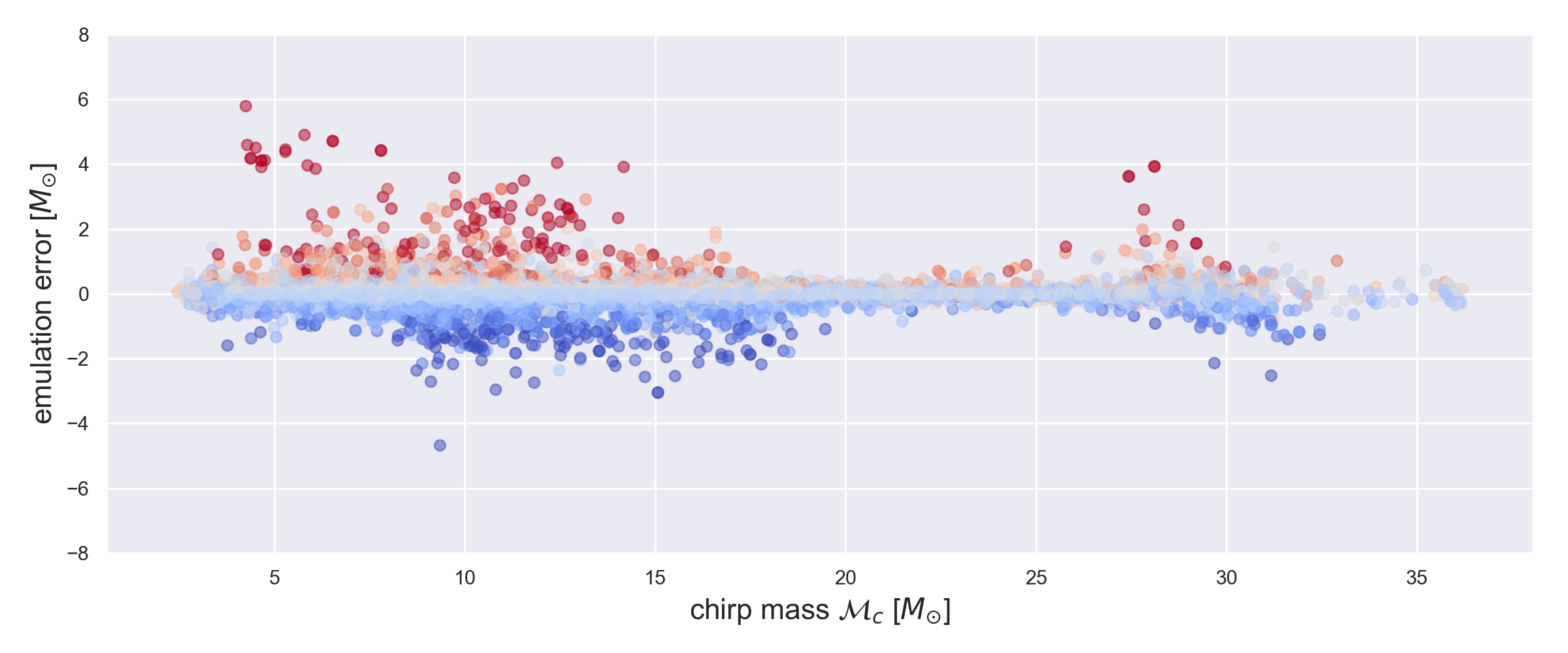}
\includegraphics[scale=.45]{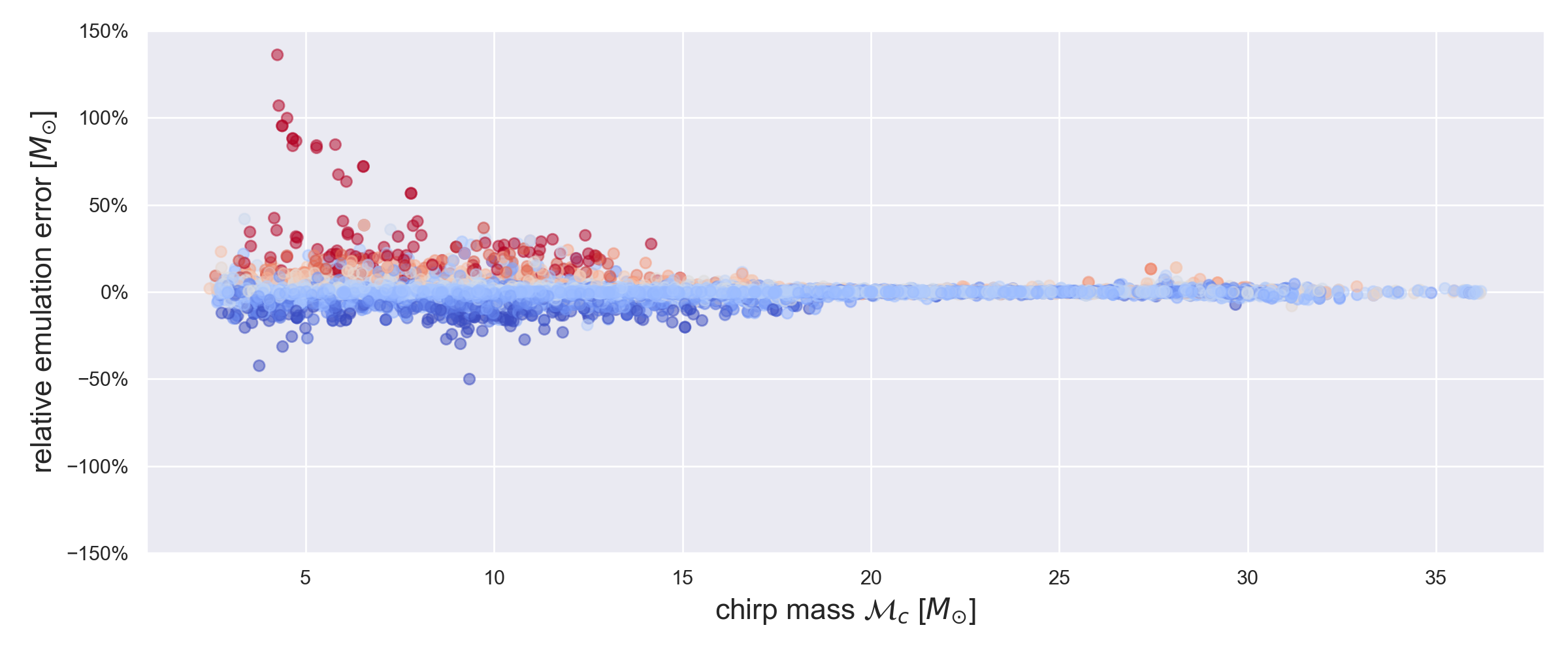}
\caption{Emulation error (upper) and relative (percentage) emulation error (lower) against the true chirp mass from cross validation. This plot combines the results of 100 out-of-sample tests with a holdout size of 1000. Only successful simulations that are classified correctly are included, to assess the performance of $\hat{z}(\cdot)$. Points are colored by the magnitude of the error. }\label{fig:emu_cv}
\end{figure}

For the \num{100000} input conditions in this cross validation, \num{41104} led to successful simulations (\num{58896} NAs), out of which \num{24419} are correctly classified by the proposed emulator. {Compared to the two-dimensional example in Section \ref{subsec:2D_toy}, this is a larger misclassification rate. We attribute this performance of the classifier to (1) the higher dimensional input space of the COMPAS model; (2) the complexity of population synthesis codes with very low success rate, and (3) input design constructed from simulating $\mbs{x}$ instead of any space-filling. } Emulation results show that for active points that are correctly classified by $\hat{y}(x)$, the Nash-Sutcliffe efficiency of all cross validations are above $99\%$, meaning that the emulator can explain more than $99\%$ of the variability in chirp mass among successful simulations that are also classified as successful by the emulator. The emulation errors (for $z(x)-\hat{z}(x)$) together with relative percentage errors (lower) are plotted in Figure \ref{fig:emu_cv} against the true chirp mass. Out of the aforementioned \num{24419} successful classifications, only $107$ cases show absolute error above $2 M_{\odot}$. A closer investigation into cases with large absolute errors revealed that the errors were caused by the sparsity of the local design $\mbs{X}^{ab}$. This issue can be addressed by applying the sequential design procedure in planning future simulations.

\subsection{Estimating the chirp mass distribution through emulation}\label{subsec:chirpMass_dist}

\begin{figure}
    \centering
    \includegraphics[scale=.42]{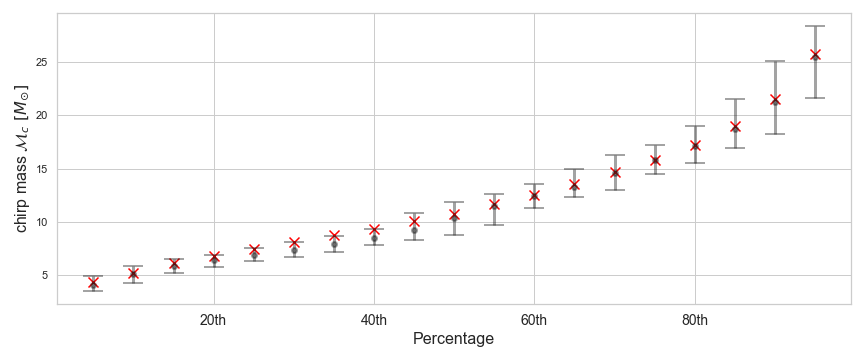}
    \caption{Confidence intervals (grey lines) and mean (grey dots) of percentiles for the chirp mass distribution, emulated with the lcGP method. Red crosses depict the corresponding percentile from an independent set of simulations \citep{broekgaarden2019stroopwafel}.}
    \label{fig:boxplot_chirpmass}
\end{figure}

We have so far treated the model output $m(x)$ as a scalar -- the chirp mass of merging binary black holes. In reality, however, two further complications arise in this context:
\begin{enumerate}[(i)]
    \item the initial conditions $x$ of a BBH cannot be observed, and are only known up to a distribution;
    \item we may be interested in emulating other properties of the merging binaries that are predicted by COMPAS population synthesis models, such as the mass ratios $m_{f,2}/m_{f,1}$ or the delay time between star formation and merger.
\end{enumerate}

We therefore consider reconstructing the distribution of $m(x)$ given the distribution of $x$, rather than emulating $m(x)$ for a given input $x$. That is, we aim to propagate the uncertainty in $x$ through the emulator $\hat{m}(x)$.  We assume that $x$ follows an initial conditions distribution $\pi(x)$, and evaluate the resulting distribution of $\hat{m}(x)$ by convolving this with the emulator.  By drawing random samples from $\pi(x)$, and building independent lcGP models in parallel for each $x$ in the sample, the distribution of $m(x)$ can be emulated.  

Furthermore, as a proxy for a bivariate output, we expand the model output to a two-dimensional vector $(m_{f,1},m_{f,2})$, representing the vector of final masses of the two black holes at the end of a simulation.  By adopting a separable covariance structure \citep{conti2010bayesian},  lcGP can easily be extended to computer models with multivariate output.  This, in particular, allows us to model the detectability of a gravitational-wave signal by the Laser Interferometer Gravitational-wave Observatory (LIGO), which depends on both component masses \citep[e.g.,][]{fishbach2017ligo,2018MNRAS.477.4685B}.

\begin{figure}[!t]
    \centering
    \begin{subfigure}{.99\linewidth}
    \includegraphics[scale=.45]{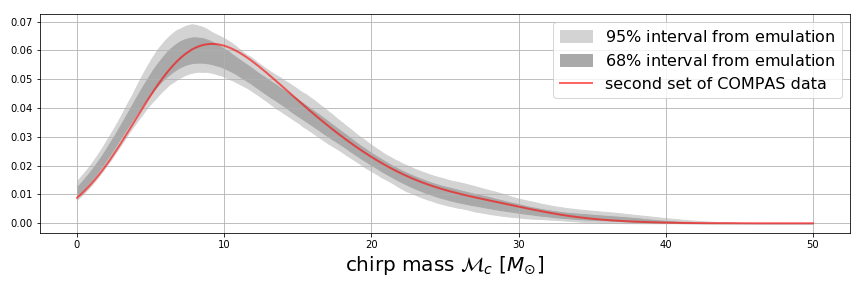}
    \caption{Kernel density estimate of the chirp mass distribution without accounting for LIGO sensitivity}\label{fig:kde_wo_ligo}
    \end{subfigure}
    \begin{subfigure}{.99\linewidth}
    \includegraphics[scale=.45]{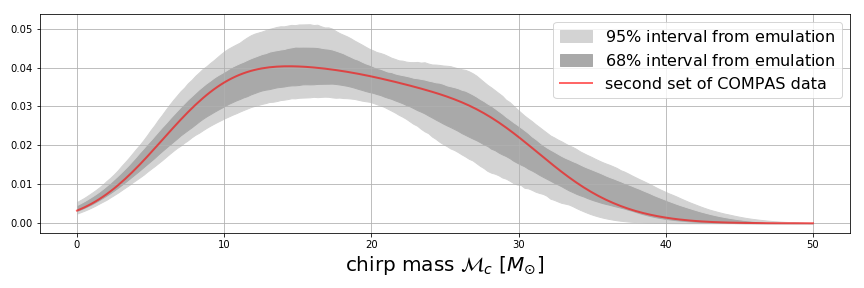}
    \caption{Kernel density estimate of the chirp mass distribution of mergers observable by LIGO}\label{fig:kde_w_ligo}
    \end{subfigure}
    \caption{Confidence band of kernel density estimate of the chirp mass obtained through emulation (grey shades), compared to the kernel density estimate of a second set of COMPAS data (solid red curve) that is independent of the modeling data. The bandwidth of the kernel is set to $0.5\ M_\odot$.}
    \label{fig:dist_chirpMass}
\end{figure}

Using the same simulation data in Section \ref{subsec:COMPAS_emu} as input, the distribution of chirp masses for the given set of population parameters $t$ is emulated with the above techniques. In Figures \ref{fig:boxplot_chirpmass} and \ref{fig:dist_chirpMass}, the emulated distribution is compared against the chirp mass distribution from an independent set of simulations conducted by sampling $x\sim\pi(x)$ in \citet{broekgaarden2019stroopwafel}\footnote{This is the COMPAS data produced with the traditional sampling method in \citet{broekgaarden2019stroopwafel}, labeled as 'Traditional'. It contains $1,000,000$ COMPAS simulations with \num{5163} successful BBHs.}.  {The percentiles (red crosses) in Figure \ref{fig:boxplot_chirpmass} and }the kernel density estimate shown with the red line in Figure \ref{fig:dist_chirpMass} is produced with the \num{5163} successful BBH mergers (out of \num{1000000} simulations) obtained from the later data set. A full Bayesian approach is taken for $\hat{m}(x)$, so that the emulated distributions take into account uncertainties 
from both GP parameters (such as $\mbs{\phi}$, $\eta$, $\mu$, $\mbs{\psi}$ and $\lambda$ discussed in Section \ref{sec:emu_model}) and the emulators ($\hat{y}(x)$ and $\hat{z}(x)$ given the GP parameters). {The same set of uniform prior distributions as in Section \ref{sec:emu_example} are chosen for the GP classifier. For length-scale parameters, $U(0,\sqrt{10})$ is used, and $U(0,4)$ is used for the inverse of precision parameters. The correlation between $m_{f,1}$ and $m_{f,2}$ is chosen to have a $U[-1,1]$ prior distribution. }The emulation procedure starts with drawing a sample of size \num{25000} from $\pi(x)$, denoted by $\mathcal{X}_{r}$, to represent the distribution of $x$. Next, for each $x_j\in\mathcal{X}_{r}$, $j=1,2,\ldots,25000$, we draw a single sample of the emulation output $\hat{m}(x_j)$. As a result, a sample for $\hat{m}(\mathcal{X}_{r})$ of size \num{25000} is obtained. 

In practice, only positive/successful outputs (BBH formations) can be detected through gravitational waves. Therefore only positive values of the above samples are kept to produce {the percentiles and } a kernel density estimate to emulate the distribution of $m(x)$. By independently repeating the emulation and kernel density estimation steps, {confidence intervals ($95\%$ confidence level) of percentiles for the emulated chirp mass distribution are obtained (Figure \ref{fig:boxplot_chirpmass}), and compared against corresponding percentiles from the independent simulation (red crosses). It can be seen that, with the percentiles considered,  the emulated confidence intervals always contain the corresponding percentile from the independent simulation. } Confidence bands of the kernel density of $\hat{m}(x)$ are generated and shown in Figure \ref{fig:dist_chirpMass}. The grey areas in Figure \ref{fig:kde_wo_ligo} indicate the $68\%$ (dark) and $95\%$ (light) confidence bands for the emulated kernel density.  The `observable' distribution of the chirp mass shown in Figure \ref{fig:kde_w_ligo} takes LIGO sensitivity into account (we approximate the LIGO sensitivity to be proportional to $\max(m_{f,1},m_{f,2})^{2.2}$ in this example). In both plots, the $95\%$ confidence band of the emulated kernel density encloses that of the independent set of COMPAS data (solid red curve).  

\section{Discussion}

In this paper, a new approach for fast emulation of computer models with unknown constraints was proposed.   When emulating the COMPAS model of BBH mergers, the proposed method enables efficient emulation through parallel computation while providing uncertainty quantification. The new sequential design criterion guides the selection of future runs to improve the exploration of the input space. By propagating the randomness of initial conditions of binary systems, the probability distribution of the BBH chirp mass was obtained. In future work, we aim to
\begin{enumerate}[(i)]
    \item {incorporate the sequential design criterion into COMPAS to improve the planning of simulations, and}
    \item compare the emulated distribution of the chirp mass at various parameter settings (e.g., $Z$, $\alpha$, $\sigma$ and flbv in Table \ref{tab:data}) to observations from LIGO/Virgo, and infer the true value of these physical parameters (i.e., {\it computer model calibration}).
\end{enumerate}

It is worth noting that the computational resources needed by the proposed emulator are largely driven by the size of the local designs in Algorithms \ref{ag:y} and \ref{ag:z}. The trade-off between computational efficiency and prediction accuracy should be carefully examined to choose $n$ and $n^M$. In general, the local emulator is more accurate when a larger local design is adopted, i.e., more information is available for emulating $m(x^*)$.  Of course, in the presence of unknown constraints, making $n^M$ arbitrarily large for example, may result in the inclusion of many points from a different constraint region, thereby decreasing the quality of the inference.  We suggest performing a preliminary analysis to find a suitable local design size for the desired accuracy and computational efficiency.

\begin{appendix}
\section{MCMC procedure to sample GP parameters}\label{app:MCMC}
Algorithm \ref{ag:MCMC_ESS} describes in detail the MCMC procedure that is applied to collect posterior samples for the GP classifier introduced in Equation \eqref{eq:GPC_posterior}.
\begin{algorithm}[!h]
 \KwIn{input design $\mbs{X}$, class label $\mbs{y}$, number of MCMC steps $n_{\textrm{MCMC}}$, initial value for $\mbs{\phi}$ and $\eta$ denoted by $\mbs{\phi}^{(0)}$ and $\eta^{(0)}$.} 
 \KwOut{posterior sample of $\mbs{f}$, $\mbs{\phi}$ and $\eta$}
 $i \leftarrow 0$\;
 $\mbs{f}^{(0)} \leftarrow \mbs{0}$\;
 $l^{(0)} \leftarrow$ log-likelihood of $(\mbs{f}^{(0)},\mbs{\phi}^{(0)},\eta^{(0)})$\;
\While{$i < n_\textrm{MCMC}$}{
    $i = i + 1$\;
    \For{$j = 1 \rightarrow d$}{
        Update $\phi_j^{(i)}$ with a Metropolis-Hastings step\;
        }
    Update $\eta_j^{(i)}$ with a Metropolis-Hastings step\;
    Update log-likelihood $l^{(i)}$ and covariance matrix $
    \Sigma^{(i)}$ with $(\mbs{f}^{(i-1)},\mbs{\phi}^{(i)},\eta^{(i)})$\;
    Sample $\mbs{\nu}\sim \mathcal{N}(0,\Sigma^{(i)})$ and $u\sim \textrm{Uniform}(0,1)$\;
    Set the acceptance threshold $\tau = l^{(i)} + \log(u)$\;
    Sample a random angle $\theta \sim \textrm{Uniform}(0,2\pi)$\;
    Set $\theta_{\textrm{min}} = \theta - 2\pi$ and $\theta_{\textrm{max}}=\theta$\;
    \While{proposal $\mbs{f}^*$ is not accepted}{
        Propose $\mbs{f}^* = \mbs{f}^{(i-1)} \cos(\theta) + \mbs{\nu}\sin(\theta)$\;
        $l^* \leftarrow$ log-likelihood with $(\mbs{f}^*,\mbs{\phi}^{(i)},\eta^{(i)})$\;
        \eIf{$l^* > \tau$}{
            Accept $\mbs{f}^*$ as $\mbs{f}^{(i)}$\;}{
            \eIf{$\theta > 0$}{Set $\theta_\textrm{max} = \theta$\;}{Set $\theta_\textrm{min} = \theta$\;}
            Draw $\theta \sim \textrm{Uniform}(\theta_\textrm{min},\theta_\textrm{max})$\;
            }
        }
    }\;
\caption{MCMC procedure for GP classifier}
\label{ag:MCMC_ESS}
\end{algorithm}
\end{appendix}
%
%

\begin{acks}[Acknowledgments]
We thank Jim Barrett and Simon Stevenson for contributions to early testing data and discussions. Simulations in this paper made use of the COMPAS rapid binary population synthesis code, which is freely available at \url{http://github.com/TeamCOMPAS/COMPAS}.  The authors would like to thank the Isaac Newton Institute for Mathematical Sciences, for support and hospitality during the program ``Uncertainty quantification for complex systems: theory and methodologies" and also the Statistical and Applied Mathematical Sciences Institute's program on Statistical, Mathematical and Computational Methods for Astronomy where work on this paper was initially undertaken.
\end{acks}
\begin{funding}
IM is a recipient of the Australian Research Council Future Fellowship FT190100574.  IM acknowledges support from the  Australian Research Council Centre of Excellence for Gravitational  Wave  Discovery  (OzGrav), through  project number CE17010000.  This work was also supported by EPSRC grant no. EP/R014604/1 and a Natural Sciences and Engineering Research Council of Canada Discovery Grant.
\end{funding}

\begin{supplement}
\stitle{Python scripts and Jupyter notebooks for numerical examples}
\sdescription{This set of complementary materials includes Python scripts and Jupyter notebooks that allow readers to reproduce the emulation results shown in Section \ref{sec:emu_example}. The two Jupyter notebooks illustrate the emulations performed with different methods in Sections \ref{subsec:1D} and \ref{subsec:2D_toy}, respectively, with all helper functions, data and results included and directly importable.}
\end{supplement}



\begin{thebibliography}{36}

\bibitem[\protect\citeauthoryear{{Andrews}, {Zezas} and
  {Fragos}}{2018}]{2018ApJS..237....1A}
\begin{barticle}[author]
\bauthor{\bsnm{{Andrews}},~\bfnm{Jeff~J.}\binits{J.~J.}},
  \bauthor{\bsnm{{Zezas}},~\bfnm{Andreas}\binits{A.}} \AND
  \bauthor{\bsnm{{Fragos}},~\bfnm{Tassos}\binits{T.}}
(\byear{2018}).
\btitle{{dart\_board: Binary Population Synthesis with Markov Chain Monte
  Carlo}}.
\bjournal{\apjs}
\bvolume{237}
\bpages{1}.
\bdoi{10.3847/1538-4365/aaca30}
\end{barticle}
\endbibitem

\bibitem[\protect\citeauthoryear{{Barrett} et~al.}{2017}]{2017IAUS..325...46B}
\begin{binproceedings}[author]
\bauthor{\bsnm{{Barrett}},~\bfnm{Jim~W.}\binits{J.~W.}},
  \bauthor{\bsnm{{Mandel}},~\bfnm{Ilya}\binits{I.}},
  \bauthor{\bsnm{{Neijssel}},~\bfnm{Coenraad~J.}\binits{C.~J.}},
  \bauthor{\bsnm{{Stevenson}},~\bfnm{Simon}\binits{S.}} \AND
  \bauthor{\bsnm{{Vigna-G{\'o}mez}},~\bfnm{Alejandro}\binits{A.}}
(\byear{2017}).
\btitle{{Exploring the Parameter Space of Compact Binary Population
  Synthesis}}.
In \bbooktitle{Astroinformatics}
(\beditor{\bfnm{Massimo}\binits{M.}~\bsnm{{Brescia}}},
  \beditor{\bfnm{S.~G.}\binits{S.~G.}~\bsnm{{Djorgovski}}},
  \beditor{\bfnm{Eric~D.}\binits{E.~D.}~\bsnm{{Feigelson}}},
  \beditor{\bfnm{Giuseppe}\binits{G.}~\bsnm{{Longo}}} \AND
  \beditor{\bfnm{Stefano}\binits{S.}~\bsnm{{Cavuoti}}}, eds.).
\bseries{IAU Symposium}
\bvolume{325}
\bpages{46-50}.
\bdoi{10.1017/S1743921317000059}
\end{binproceedings}
\endbibitem

\bibitem[\protect\citeauthoryear{{Barrett} et~al.}{2018}]{2018MNRAS.477.4685B}
\begin{barticle}[author]
\bauthor{\bsnm{{Barrett}},~\bfnm{J.~W.}\binits{J.~W.}},
  \bauthor{\bsnm{{Gaebel}},~\bfnm{S.~M.}\binits{S.~M.}},
  \bauthor{\bsnm{{Neijssel}},~\bfnm{C.~J.}\binits{C.~J.}},
  \bauthor{\bsnm{{Vigna-G{\'o}mez}},~\bfnm{A.}\binits{A.}},
  \bauthor{\bsnm{{Stevenson}},~\bfnm{S.}\binits{S.}},
  \bauthor{\bsnm{{Berry}},~\bfnm{C.~P.~L.}\binits{C.~P.~L.}},
  \bauthor{\bsnm{{Farr}},~\bfnm{W.~M.}\binits{W.~M.}} \AND
  \bauthor{\bsnm{{Mandel}},~\bfnm{I.}\binits{I.}}
(\byear{2018}).
\btitle{{Accuracy of inference on the physics of binary evolution from
  gravitational-wave observations}}.
\bjournal{\mnras}
\bvolume{477}
\bpages{4685-4695}.
\bdoi{10.1093/mnras/sty908}
\end{barticle}
\endbibitem

\bibitem[\protect\citeauthoryear{{Belczynski}, {Kalogera} and
  {Bulik}}{2002}]{2002ApJ...572..407B}
\begin{barticle}[author]
\bauthor{\bsnm{{Belczynski}},~\bfnm{Krzysztof}\binits{K.}},
  \bauthor{\bsnm{{Kalogera}},~\bfnm{Vassiliki}\binits{V.}} \AND
  \bauthor{\bsnm{{Bulik}},~\bfnm{Tomasz}\binits{T.}}
(\byear{2002}).
\btitle{{A Comprehensive Study of Binary Compact Objects as Gravitational Wave
  Sources: Evolutionary Channels, Rates, and Physical Properties}}.
\bjournal{\apj}
\bvolume{572}
\bpages{407-431}.
\bdoi{10.1086/340304}
\end{barticle}
\endbibitem

\bibitem[\protect\citeauthoryear{Bingham, Ranjan and
  Welch}{2014}]{bingham2014design}
\begin{barticle}[author]
\bauthor{\bsnm{Bingham},~\bfnm{Derek}\binits{D.}},
  \bauthor{\bsnm{Ranjan},~\bfnm{Pritam}\binits{P.}} \AND
  \bauthor{\bsnm{Welch},~\bfnm{William~J}\binits{W.~J.}}
(\byear{2014}).
\btitle{Design of computer experiments for optimization, estimation of function
  contours, and related objectives}.
\bjournal{Statistics in Action: A Canadian Outlook}
\bvolume{109}.
\end{barticle}
\endbibitem

\bibitem[\protect\citeauthoryear{{Broekgaarden}
  et~al.}{2019}]{broekgaarden2019stroopwafel}
\begin{barticle}[author]
\bauthor{\bsnm{{Broekgaarden}},~\bfnm{Floor~S.}\binits{F.~S.}},
  \bauthor{\bsnm{{Justham}},~\bfnm{Stephen}\binits{S.}}, \bauthor{\bsnm{{de
  Mink}},~\bfnm{Selma~E.}\binits{S.~E.}},
  \bauthor{\bsnm{{Gair}},~\bfnm{Jonathan}\binits{J.}},
  \bauthor{\bsnm{{Mandel}},~\bfnm{Ilya}\binits{I.}},
  \bauthor{\bsnm{{Stevenson}},~\bfnm{Simon}\binits{S.}},
  \bauthor{\bsnm{{Barrett}},~\bfnm{Jim~W.}\binits{J.~W.}},
  \bauthor{\bsnm{{Vigna-G{\'o}mez}},~\bfnm{Alejandro}\binits{A.}} \AND
  \bauthor{\bsnm{{Neijssel}},~\bfnm{Coenraad~J.}\binits{C.~J.}}
(\byear{2019}).
\btitle{{STROOPWAFEL: simulating rare outcomes from astrophysical populations,
  with application to gravitational-wave sources}}.
\bjournal{\mnras}
\bvolume{490}
\bpages{5228-5248}.
\bdoi{10.1093/mnras/stz2558}
\end{barticle}
\endbibitem

\bibitem[\protect\citeauthoryear{Conti and O'Hagan}{2010}]{conti2010bayesian}
\begin{barticle}[author]
\bauthor{\bsnm{Conti},~\bfnm{Stefano}\binits{S.}} \AND
  \bauthor{\bsnm{O'Hagan},~\bfnm{Anthony}\binits{A.}}
(\byear{2010}).
\btitle{Bayesian emulation of complex multi-output and dynamic computer
  models}.
\bjournal{Journal of statistical planning and inference}
\bvolume{140}
\bpages{640--651}.
\end{barticle}
\endbibitem

\bibitem[\protect\citeauthoryear{Cover and Hart}{1967}]{cover1967nearest}
\begin{barticle}[author]
\bauthor{\bsnm{Cover},~\bfnm{Thomas}\binits{T.}} \AND
  \bauthor{\bsnm{Hart},~\bfnm{Peter}\binits{P.}}
(\byear{1967}).
\btitle{Nearest neighbor pattern classification}.
\bjournal{IEEE transactions on information theory}
\bvolume{13}
\bpages{21--27}.
\end{barticle}
\endbibitem

\bibitem[\protect\citeauthoryear{Cressie and
  Johannesson}{2008}]{cressie2008fixed}
\begin{barticle}[author]
\bauthor{\bsnm{Cressie},~\bfnm{Noel}\binits{N.}} \AND
  \bauthor{\bsnm{Johannesson},~\bfnm{Gardar}\binits{G.}}
(\byear{2008}).
\btitle{Fixed rank kriging for very large spatial data sets}.
\bjournal{Journal of the Royal Statistical Society: Series B (Statistical
  Methodology)}
\bvolume{70}
\bpages{209--226}.
\end{barticle}
\endbibitem

\bibitem[\protect\citeauthoryear{Dunlop et~al.}{2018}]{dunlop2018deep}
\begin{barticle}[author]
\bauthor{\bsnm{Dunlop},~\bfnm{Matthew~M}\binits{M.~M.}},
  \bauthor{\bsnm{Girolami},~\bfnm{Mark~A}\binits{M.~A.}},
  \bauthor{\bsnm{Stuart},~\bfnm{Andrew~M}\binits{A.~M.}} \AND
  \bauthor{\bsnm{Teckentrup},~\bfnm{Aretha~L}\binits{A.~L.}}
(\byear{2018}).
\btitle{How deep are deep Gaussian processes?}
\bjournal{The Journal of Machine Learning Research}
\bvolume{19}
\bpages{2100--2145}.
\end{barticle}
\endbibitem

\bibitem[\protect\citeauthoryear{Fishbach and Holz}{2017}]{fishbach2017ligo}
\begin{barticle}[author]
\bauthor{\bsnm{Fishbach},~\bfnm{Maya}\binits{M.}} \AND
  \bauthor{\bsnm{Holz},~\bfnm{Daniel~E}\binits{D.~E.}}
(\byear{2017}).
\btitle{Where are LIGO's big black holes?}
\bjournal{The Astrophysical Journal Letters}
\bvolume{851}
\bpages{L25}.
\end{barticle}
\endbibitem

\bibitem[\protect\citeauthoryear{Gelbart, Snoek and
  Adams}{2014}]{gelbart2014bayesian}
\begin{barticle}[author]
\bauthor{\bsnm{Gelbart},~\bfnm{Michael~A}\binits{M.~A.}},
  \bauthor{\bsnm{Snoek},~\bfnm{Jasper}\binits{J.}} \AND
  \bauthor{\bsnm{Adams},~\bfnm{Ryan~P}\binits{R.~P.}}
(\byear{2014}).
\btitle{Bayesian optimization with unknown constraints}.
\bjournal{arXiv preprint arXiv:1403.5607}.
\end{barticle}
\endbibitem

\bibitem[\protect\citeauthoryear{Gramacy and Apley}{2015}]{gramacy2015local}
\begin{barticle}[author]
\bauthor{\bsnm{Gramacy},~\bfnm{Robert~B}\binits{R.~B.}} \AND
  \bauthor{\bsnm{Apley},~\bfnm{Daniel~W}\binits{D.~W.}}
(\byear{2015}).
\btitle{Local Gaussian process approximation for large computer experiments}.
\bjournal{Journal of Computational and Graphical Statistics}
\bvolume{24}
\bpages{561--578}.
\end{barticle}
\endbibitem

\bibitem[\protect\citeauthoryear{Gramacy and Lee}{2008}]{gramacy2008bayesian}
\begin{barticle}[author]
\bauthor{\bsnm{Gramacy},~\bfnm{Robert~B}\binits{R.~B.}} \AND
  \bauthor{\bsnm{Lee},~\bfnm{Herbert K~H}\binits{H.~K.~H.}}
(\byear{2008}).
\btitle{Bayesian treed Gaussian process models with an application to computer
  modeling}.
\bjournal{Journal of the American Statistical Association}
\bvolume{103}
\bpages{1119--1130}.
\end{barticle}
\endbibitem

\bibitem[\protect\citeauthoryear{Gramacy and
  Lee}{2010}]{gramacy2010optimization}
\begin{bmisc}[author]
\bauthor{\bsnm{Gramacy},~\bfnm{Robert~B.}\binits{R.~B.}} \AND
  \bauthor{\bsnm{Lee},~\bfnm{Herbert K.~H.}\binits{H.~K.~H.}}
(\byear{2010}).
\btitle{Optimization Under Unknown Constraints}.
\end{bmisc}
\endbibitem

\bibitem[\protect\citeauthoryear{Hastings}{1970}]{hastings1970monte}
\begin{barticle}[author]
\bauthor{\bsnm{Hastings},~\bfnm{W~Keith}\binits{W.~K.}}
(\byear{1970}).
\btitle{Monte Carlo sampling methods using Markov chains and their
  applications}.
\end{barticle}
\endbibitem

\bibitem[\protect\citeauthoryear{Higdon et~al.}{2008}]{higdon2008computer}
\begin{barticle}[author]
\bauthor{\bsnm{Higdon},~\bfnm{Dave}\binits{D.}},
  \bauthor{\bsnm{Gattiker},~\bfnm{James}\binits{J.}},
  \bauthor{\bsnm{Williams},~\bfnm{Brian}\binits{B.}} \AND
  \bauthor{\bsnm{Rightley},~\bfnm{Maria}\binits{M.}}
(\byear{2008}).
\btitle{Computer model calibration using high-dimensional output}.
\bjournal{Journal of the American Statistical Association}
\bvolume{103}
\bpages{570--583}.
\end{barticle}
\endbibitem

\bibitem[\protect\citeauthoryear{Hsu}{2019}]{hsu2019fast}
\begin{barticle}[author]
\bauthor{\bsnm{Hsu},~\bfnm{Grace}\binits{G.}}
(\byear{2019}).
\btitle{Fast emulation and calibration of large computer experiments with
  multivariate output}.
\bjournal{Unpublished M.Sc. Thesis, Department of Statistics and Actuarial
  Science, Simon Fraser University}.
\end{barticle}
\endbibitem

\bibitem[\protect\citeauthoryear{Jones, Schonlau and
  Welch}{1998}]{jones1998efficient}
\begin{barticle}[author]
\bauthor{\bsnm{Jones},~\bfnm{Donald~R}\binits{D.~R.}},
  \bauthor{\bsnm{Schonlau},~\bfnm{Matthias}\binits{M.}} \AND
  \bauthor{\bsnm{Welch},~\bfnm{William~J}\binits{W.~J.}}
(\byear{1998}).
\btitle{Efficient global optimization of expensive black-box functions}.
\bjournal{Journal of Global optimization}
\bvolume{13}
\bpages{455--492}.
\end{barticle}
\endbibitem

\bibitem[\protect\citeauthoryear{Kaufman et~al.}{2011}]{kaufman2011efficient}
\begin{barticle}[author]
\bauthor{\bsnm{Kaufman},~\bfnm{Cari~G}\binits{C.~G.}},
  \bauthor{\bsnm{Bingham},~\bfnm{Derek}\binits{D.}},
  \bauthor{\bsnm{Habib},~\bfnm{Salman}\binits{S.}},
  \bauthor{\bsnm{Heitmann},~\bfnm{Katrin}\binits{K.}},
  \bauthor{\bsnm{Frieman},~\bfnm{Joshua~A}\binits{J.~A.}} \betal{et~al.}
(\byear{2011}).
\btitle{Efficient emulators of computer experiments using compactly supported
  correlation functions, with an application to cosmology}.
\bjournal{The Annals of Applied Statistics}
\bvolume{5}
\bpages{2470--2492}.
\end{barticle}
\endbibitem

\bibitem[\protect\citeauthoryear{{Kruckow} et~al.}{2018}]{2018MNRAS.481.1908K}
\begin{barticle}[author]
\bauthor{\bsnm{{Kruckow}},~\bfnm{Matthias~U.}\binits{M.~U.}},
  \bauthor{\bsnm{{Tauris}},~\bfnm{Thomas~M.}\binits{T.~M.}},
  \bauthor{\bsnm{{Langer}},~\bfnm{Norbert}\binits{N.}},
  \bauthor{\bsnm{{Kramer}},~\bfnm{Michael}\binits{M.}} \AND
  \bauthor{\bsnm{{Izzard}},~\bfnm{Robert~G.}\binits{R.~G.}}
(\byear{2018}).
\btitle{{Progenitors of gravitational wave mergers: binary evolution with the
  stellar grid-based code COMBINE}}.
\bjournal{\mnras}
\bvolume{481}
\bpages{1908-1949}.
\bdoi{10.1093/mnras/sty2190}
\end{barticle}
\endbibitem

\bibitem[\protect\citeauthoryear{Lawrence et~al.}{2017}]{lawrence2017mira}
\begin{barticle}[author]
\bauthor{\bsnm{Lawrence},~\bfnm{Earl}\binits{E.}},
  \bauthor{\bsnm{Heitmann},~\bfnm{Katrin}\binits{K.}},
  \bauthor{\bsnm{Kwan},~\bfnm{Juliana}\binits{J.}},
  \bauthor{\bsnm{Upadhye},~\bfnm{Amol}\binits{A.}},
  \bauthor{\bsnm{Bingham},~\bfnm{Derek}\binits{D.}},
  \bauthor{\bsnm{Habib},~\bfnm{Salman}\binits{S.}},
  \bauthor{\bsnm{Higdon},~\bfnm{David}\binits{D.}},
  \bauthor{\bsnm{Pope},~\bfnm{Adrian}\binits{A.}},
  \bauthor{\bsnm{Finkel},~\bfnm{Hal}\binits{H.}} \AND
  \bauthor{\bsnm{Frontiere},~\bfnm{Nicholas}\binits{N.}}
(\byear{2017}).
\btitle{The Mira-Titan universe. II. Matter power spectrum emulation}.
\bjournal{The Astrophysical Journal}
\bvolume{847}
\bpages{50}.
\end{barticle}
\endbibitem

\bibitem[\protect\citeauthoryear{Lin et~al.}{2021}]{supplement}
\begin{barticle}[author]
\bauthor{\bsnm{Lin},~\bfnm{Luyao}\binits{L.}},
  \bauthor{\bsnm{Bingham},~\bfnm{Derek}\binits{D.}},
  \bauthor{\bsnm{Broekgaarden},~\bfnm{Floor}\binits{F.}} \AND
  \bauthor{\bsnm{Mandel},~\bfnm{Ilya}\binits{I.}}
(\byear{2021}).
\btitle{{Supplement to ``Uncertainty quantification of a computer model for
  binary black hole formation''}}.
\bdoi{10.1214/[provided by typesetter]}
\end{barticle}
\endbibitem

\bibitem[\protect\citeauthoryear{Mandel and
  Farmer}{2017}]{Mandel2017GravitationalWS}
\begin{barticle}[author]
\bauthor{\bsnm{Mandel},~\bfnm{Ilya}\binits{I.}} \AND
  \bauthor{\bsnm{Farmer},~\bfnm{Alison}\binits{A.}}
(\byear{2017}).
\btitle{Gravitational waves: Stellar palaeontology}.
\bjournal{Nature}
\bvolume{547}
\bpages{284-285}.
\end{barticle}
\endbibitem

\bibitem[\protect\citeauthoryear{{Mandel} and
  {Farmer}}{2018}]{MandelFarmer:2018}
\begin{barticle}[author]
\bauthor{\bsnm{{Mandel}},~\bfnm{I.}\binits{I.}} \AND
  \bauthor{\bsnm{{Farmer}},~\bfnm{A.}\binits{A.}}
(\byear{2018}).
\btitle{{Merging stellar-mass binary black holes}}.
\bjournal{ArXiv e-prints}.
\end{barticle}
\endbibitem

\bibitem[\protect\citeauthoryear{Murray, Prescott~Adams and
  MacKay}{2010}]{murray2010elliptical}
\begin{barticle}[author]
\bauthor{\bsnm{Murray},~\bfnm{Iain}\binits{I.}},
  \bauthor{\bsnm{Prescott~Adams},~\bfnm{Ryan}\binits{R.}} \AND
  \bauthor{\bsnm{MacKay},~\bfnm{David~JC}\binits{D.~J.}}
(\byear{2010}).
\btitle{Elliptical slice sampling}.
\end{barticle}
\endbibitem

\bibitem[\protect\citeauthoryear{Nash and Sutcliffe}{1970}]{NASH1970282}
\begin{barticle}[author]
\bauthor{\bsnm{Nash},~\bfnm{J.~E.}\binits{J.~E.}} \AND
  \bauthor{\bsnm{Sutcliffe},~\bfnm{J.~V.}\binits{J.~V.}}
(\byear{1970}).
\btitle{River flow forecasting through conceptual models part I — A
  discussion of principles}.
\bjournal{Journal of Hydrology}
\bvolume{10}
\bpages{282 - 290}.
\bdoi{https://doi.org/10.1016/0022-1694(70)90255-6}
\end{barticle}
\endbibitem

\bibitem[\protect\citeauthoryear{{Neijssel} et~al.}{2019}]{Neijssel:2019}
\begin{barticle}[author]
\bauthor{\bsnm{{Neijssel}},~\bfnm{Coenraad~J.}\binits{C.~J.}},
  \bauthor{\bsnm{{Vigna-G{\'o}mez}},~\bfnm{Alejandro}\binits{A.}},
  \bauthor{\bsnm{{Stevenson}},~\bfnm{Simon}\binits{S.}},
  \bauthor{\bsnm{{Barrett}},~\bfnm{Jim~W.}\binits{J.~W.}},
  \bauthor{\bsnm{{Gaebel}},~\bfnm{Sebastian~M.}\binits{S.~M.}},
  \bauthor{\bsnm{{Broekgaarden}},~\bfnm{Floor~S.}\binits{F.~S.}},
  \bauthor{\bsnm{{de Mink}},~\bfnm{Selma~E.}\binits{S.~E.}},
  \bauthor{\bsnm{{Sz{\'e}csi}},~\bfnm{Dorottya}\binits{D.}},
  \bauthor{\bsnm{{Vinciguerra}},~\bfnm{Serena}\binits{S.}} \AND
  \bauthor{\bsnm{{Mandel}},~\bfnm{Ilya}\binits{I.}}
(\byear{2019}).
\btitle{{The effect of the metallicity-specific star formation history on
  double compact object mergers}}.
\bjournal{\mnras}
\bvolume{490}
\bpages{3740-3759}.
\bdoi{10.1093/mnras/stz2840}
\end{barticle}
\endbibitem

\bibitem[\protect\citeauthoryear{{Peters} and
  {Mathews}}{1963}]{1963PhRv..131..435P}
\begin{barticle}[author]
\bauthor{\bsnm{{Peters}},~\bfnm{P.~C.}\binits{P.~C.}} \AND
  \bauthor{\bsnm{{Mathews}},~\bfnm{J.}\binits{J.}}
(\byear{1963}).
\btitle{{Gravitational Radiation from Point Masses in a Keplerian Orbit}}.
\bjournal{Physical Review}
\bvolume{131}
\bpages{435-440}.
\bdoi{10.1103/PhysRev.131.435}
\end{barticle}
\endbibitem

\bibitem[\protect\citeauthoryear{Qui{\~n}onero-Candela and
  Rasmussen}{2005}]{quinonero2005unifying}
\begin{barticle}[author]
\bauthor{\bsnm{Qui{\~n}onero-Candela},~\bfnm{Joaquin}\binits{J.}} \AND
  \bauthor{\bsnm{Rasmussen},~\bfnm{Carl~Edward}\binits{C.~E.}}
(\byear{2005}).
\btitle{A unifying view of sparse approximate Gaussian process regression}.
\bjournal{Journal of Machine Learning Research}
\bvolume{6}
\bpages{1939--1959}.
\end{barticle}
\endbibitem

\bibitem[\protect\citeauthoryear{Sacks et~al.}{1989}]{sacks1989design}
\begin{barticle}[author]
\bauthor{\bsnm{Sacks},~\bfnm{Jerome}\binits{J.}},
  \bauthor{\bsnm{Welch},~\bfnm{William~J}\binits{W.~J.}},
  \bauthor{\bsnm{Mitchell},~\bfnm{Toby~J}\binits{T.~J.}} \AND
  \bauthor{\bsnm{Wynn},~\bfnm{Henry~P}\binits{H.~P.}}
(\byear{1989}).
\btitle{Design and analysis of computer experiments}.
\bjournal{Statistical science}
\bpages{409--423}.
\end{barticle}
\endbibitem

\bibitem[\protect\citeauthoryear{Stevenson
  et~al.}{2017}]{stevenson2017formation}
\begin{barticle}[author]
\bauthor{\bsnm{Stevenson},~\bfnm{Simon}\binits{S.}},
  \bauthor{\bsnm{Vigna-G{\'o}mez},~\bfnm{Alejandro}\binits{A.}},
  \bauthor{\bsnm{Mandel},~\bfnm{Ilya}\binits{I.}},
  \bauthor{\bsnm{Barrett},~\bfnm{Jim~W}\binits{J.~W.}},
  \bauthor{\bsnm{Neijssel},~\bfnm{Coenraad~J}\binits{C.~J.}},
  \bauthor{\bsnm{Perkins},~\bfnm{David}\binits{D.}} \AND
  \bauthor{\bparticle{de} \bsnm{Mink},~\bfnm{Selma~E}\binits{S.~E.}}
(\byear{2017}).
\btitle{Formation of the first three gravitational-wave observations through
  isolated binary evolution}.
\bjournal{Nature Communications}
\bvolume{8}.
\end{barticle}
\endbibitem

\bibitem[\protect\citeauthoryear{Tang}{1993}]{tang1993orthogonal}
\begin{barticle}[author]
\bauthor{\bsnm{Tang},~\bfnm{Boxin}\binits{B.}}
(\byear{1993}).
\btitle{Orthogonal array-based Latin hypercubes}.
\bjournal{Journal of the American statistical association}
\bvolume{88}
\bpages{1392--1397}.
\end{barticle}
\endbibitem

\bibitem[\protect\citeauthoryear{{Taylor} and
  {Gerosa}}{2018}]{2018PhRvD..98h3017T}
\begin{barticle}[author]
\bauthor{\bsnm{{Taylor}},~\bfnm{Stephen~R.}\binits{S.~R.}} \AND
  \bauthor{\bsnm{{Gerosa}},~\bfnm{Davide}\binits{D.}}
(\byear{2018}).
\btitle{{Mining gravitational-wave catalogs to understand binary stellar
  evolution: A new hierarchical Bayesian framework}}.
\bjournal{\prd}
\bvolume{98}
\bpages{083017}.
\bdoi{10.1103/PhysRevD.98.083017}
\end{barticle}
\endbibitem

\bibitem[\protect\citeauthoryear{{Vigna-G{\'o}mez}
  et~al.}{2018}]{2018MNRAS.481.4009V}
\begin{barticle}[author]
\bauthor{\bsnm{{Vigna-G{\'o}mez}},~\bfnm{Alejandro}\binits{A.}},
  \bauthor{\bsnm{{Neijssel}},~\bfnm{Coenraad~J.}\binits{C.~J.}},
  \bauthor{\bsnm{{Stevenson}},~\bfnm{Simon}\binits{S.}},
  \bauthor{\bsnm{{Barrett}},~\bfnm{Jim~W.}\binits{J.~W.}},
  \bauthor{\bsnm{{Belczynski}},~\bfnm{Krzysztof}\binits{K.}},
  \bauthor{\bsnm{{Justham}},~\bfnm{Stephen}\binits{S.}}, \bauthor{\bsnm{{de
  Mink}},~\bfnm{Selma~E.}\binits{S.~E.}},
  \bauthor{\bsnm{{M{\"u}ller}},~\bfnm{Bernhard}\binits{B.}},
  \bauthor{\bsnm{{Podsiadlowski}},~\bfnm{Philipp}\binits{P.}},
  \bauthor{\bsnm{{Renzo}},~\bfnm{Mathieu}\binits{M.}},
  \bauthor{\bsnm{{Sz{\'e}csi}},~\bfnm{Dorottya}\binits{D.}} \AND
  \bauthor{\bsnm{{Mandel}},~\bfnm{Ilya}\binits{I.}}
(\byear{2018}).
\btitle{{On the formation history of Galactic double neutron stars}}.
\bjournal{\mnras}
\bvolume{481}
\bpages{4009-4029}.
\bdoi{10.1093/mnras/sty2463}
\end{barticle}
\endbibitem

\bibitem[\protect\citeauthoryear{Williams and
  Rasmussen}{2006}]{williams2006gaussian}
\begin{bbook}[author]
\bauthor{\bsnm{Williams},~\bfnm{Christopher~KI}\binits{C.~K.}} \AND
  \bauthor{\bsnm{Rasmussen},~\bfnm{Carl~Edward}\binits{C.~E.}}
(\byear{2006}).
\btitle{Gaussian processes for machine learning}
\bvolume{2}.
\bpublisher{MIT press Cambridge, MA}.
\end{bbook}
\endbibitem

\end{thebibliography}
\end{document}